\documentclass[slac_one]{revtex4}

\usepackage{color}

\usepackage{graphicx}

\usepackage{fancyhdr}

\definecolor{Black}{named}{Black}
\definecolor{Blue}{named}{Blue}
\definecolor{Red}{named}{Red}
\definecolor{Green}{named}{ForestGreen}
\definecolor{Black}{named}{Black}
\definecolor{Olive}{named}{OliveGreen}
\definecolor{Royal}{named}{RoyalBlue}
\definecolor{Orange}{named}{YellowOrange}
\definecolor{Yellow}{named}{Goldenrod}
\definecolor{Cornblue}{named}{CornflowerBlue}
\definecolor{Lila}{named}{DarkOrchid}

\newcommand{\GeV} {\mathrm{GeV}}
\newcommand{\cL } {{\cal L}}
\newcommand{\fbi} {\mathrm{fb}^{-1}}

\def\tq{{\tilde q}}
\def\tchi{{\tilde\chi}}
\def\tl{{\tilde\ell}}
\def\lsp{{\tilde\chi_1^0}}

\def\tell{{\tilde\ell}}

\def\stau  {{\tilde\tau}}

\def\mq{m_{\tilde{q}_L}^2}
\def\mT{m_{\tilde{\chi}_2^0}^2}
\def\ml{m_{\tilde{l}_R}^2}
\def\mlfour{m_{\tilde{l}_R}^4}
\def\mO{m_{\tilde{\chi}_1^0}^2}
\def\threshold{{\rm thres}}
\def\edge{{\rm edge}}
\def\max{{\rm max}}
\def\min{{\rm min}}
\def\MT2{M_{T2}}

\begin{document}


\title{Supersymmetry at LHC and ILC\footnote{to appear in Proceedings
of the SLAC Summer Institute 2004}}


%




\author{K. Desch}

\affiliation{University of Hamburg, 22761 Hamburg, Luruper Chaussee 149, Germany\\ now at University of Freiburg, 79104 Freiburg, Hermann-Herder-Str.3, Germany}

\begin{abstract}

The prospects for the discovery and exploration of low-energy
Supersymmetry at future colliders, the Large Hadron Collider (LHC)
and the future international linear electron positron collider (ILC) 
are summarized. The focus is on the experimental techniques that will
be used to discover superpartners and to measure their properties.
Special attention is given to the question how the results from
both machines could influence each other, in particular when they
have overlapping running time.
\end{abstract}


\maketitle

\thispagestyle{fancy}





\section{INTRODUCTION}

The search for SUSY and, should it be 
found, measurements of the superpartner properties are among the 
most important motivations for future high energy particle colliders. 
A general introduction to TeV-scale Supersymmetry (SUSY) as one of the best
motivated extensions of the Standard Model (SM) has been given elsewhere
in these proceedings~\cite{tata}. 
The Large Hadron Collider (LHC) currently under construction at CERN
will go into operation in 2007 and has a huge potential for SUSY discovery
as well as for first measurements of SUSY particle properties. The planned
International Linear electron positron Collider (ILC) is an ideal tool
for precision SUSY measurements. Both machines together will be able to
give important insight into the mechanism of SUSY breaking and may open a 
window to GUT/Planck scale physics.

In this article, in Section~\ref{sec:lhc} the prospects for
inclusive SUSY discovery at the LHC are summarized. Furthermore
studies for the exclusive reconstruction of superpartners and the
measurement of their masses are explained. In Section~\ref{sec:lc}
the prospects for precision SUSY measurements at the ILC are shown. The 
different techniques for mass measurements, measurements of polarized 
cross-sections and quantum numbers of the superpartners are detailed
with some explicit examples. Finally, in Section~\ref{sec:lhclc} the 
interplay of the anticipated results of both LHC and and ILC in 
particular when analyzed simultaneously is shown. The extended
Higgs-sector of supersymmetric models also is an integral part of
future exploration of SUSY. The prospects for Higgs searches and 
precision measurements at LHC and ILC are summarized elsewhere in 
these proceedings~\cite{albert}.

\begin{figure*}[t]
\centering
\includegraphics[width=10cm,clip]{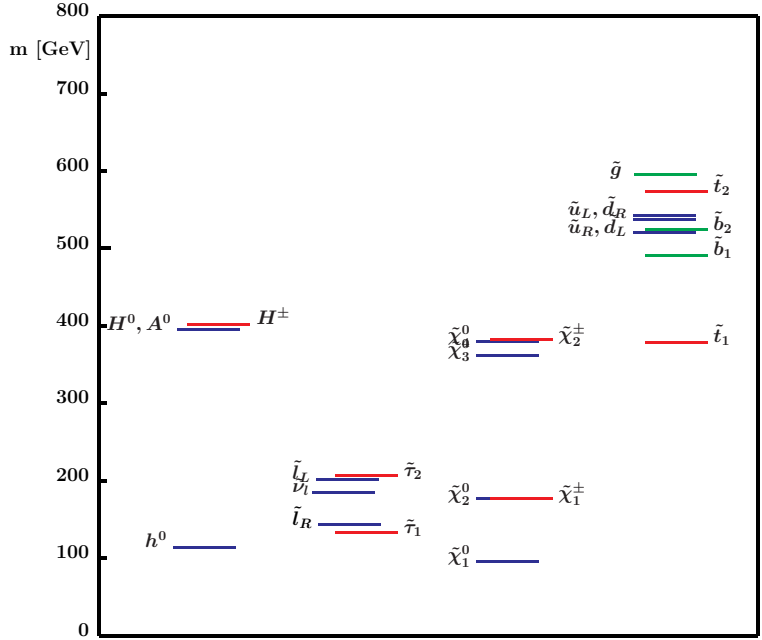}
\caption{SUSY mass spectrum for the mSUGRA benchmark point SPS1a~\cite{sps1a}.}
\label{fig:sps1a}
\end{figure*}

\section{SUSY AT THE LHC}
\label{sec:lhc}

\subsection{Experimental environment}

The LHC will produce proton-proton collisions at a center-of-mass energy
of 14~TeV initially at a luminosity 
of $1\times 10^{33} {\rm cm}^{-2}{\rm s}^{-1}$ 
(low-luminosity) and later at 
$1\times 10^{34} {\rm cm}^{-2}{\rm s}^{-1}$ (high luminosity).
The high center-of-mass energy makes this machine well suited for the
direct discovery of new massive particles beyond the SM. 
The partonic luminosity is large enough to pair-produce colored new particles
(like squarks and gluinos) up to masses of a few TeV at observable
rates. The high luminosity
and the large total pp cross-section impose strong requirements on the 
performance of the detectors and in particular on the trigger. 
The two multi-purpose detectors, ATLAS~\cite{ATLAS} 
and CMS~\cite{CMS}, will be able to trigger
efficiently on high-$p_T$ leptons, jets and are due to their hermeticity
sensitive to missing transverse energy. Thus they are well suited for 
the typical experimental signatures of SUSY. The hadronic environment,
however makes the exclusive reconstruction of final states as well as
precision measurements challenging. This is due to generally 
huge QCD backgrounds, the presence of pile-up events at high luminosity,
and the absence of a longitudinal beam constraint. In the past years 
several sophisticated analysis techniques for SUSY processes at the LHC
have been developed which in spite of the difficult conditions go far
beyond inclusive SUSY discovery and
will allow for a significant set of first measurements of superpartner
masses and some of their properties at least under favorable circumstances.

\subsection{Production processes}

At the LHC, the predominantly produced superpartners will be the
colored gluinos and squarks. If R-parity is conserved, they will be
pair-produced at large rates (typically $\cal{O}$(10~pb) at masses
around 1 TeV), comparable to the SM jet rates at the same
values of $Q^2 = M^2_{SUSY}$. 
Direct production of sleptons, charginos and neutralinos mainly proceeds via
Drell-Yan production and t-channel squark exchange 
at a much lower rate. However, the color-neutral
superpartners often appear in the decay chains of squarks, if kinematically
allowed.

If R-parity is conserved, the Lightest Supersymmetric Particle (LSP)
is expected to be neutral, stable and only weakly interacting. It escapes
detection and leads to the most distinctive SUSY signature: large missing
transverse energy since all superpartner decay chains eventually
end in the LSP. In mSUGRA and AMSB SUSY breaking models, 
the LSP is the lightest neutralino, in GMSB it is the gravitino. 

In Fig.~\ref{fig:sps1a} the superpartner spectrum for typical mSUGRA
point, SPS1a~\cite{sps1a} is shown. This particular benchmark scenario
has been extensively studied both for the LHC and for the ILC. It
provides a very rich phenomenology at both machines since the
complete spectrum lies below 600~GeV. For the LHC also a larger set of
post-LEP benchmark points corresponding to various SUSY breaking
mechanisms and parameter sets has been studied.

If R-parity is broken, the missing energy signature gets lost. Depending
on the R-parity violating model multi-jet and/or multi-lepton signatures
arise. They have also been studied~\cite{ATLAS} but will not be further
discussed here.

\subsection{Inclusive discovery}

\begin{figure*}[htbp]
\centering
\includegraphics[width=9cm,clip]{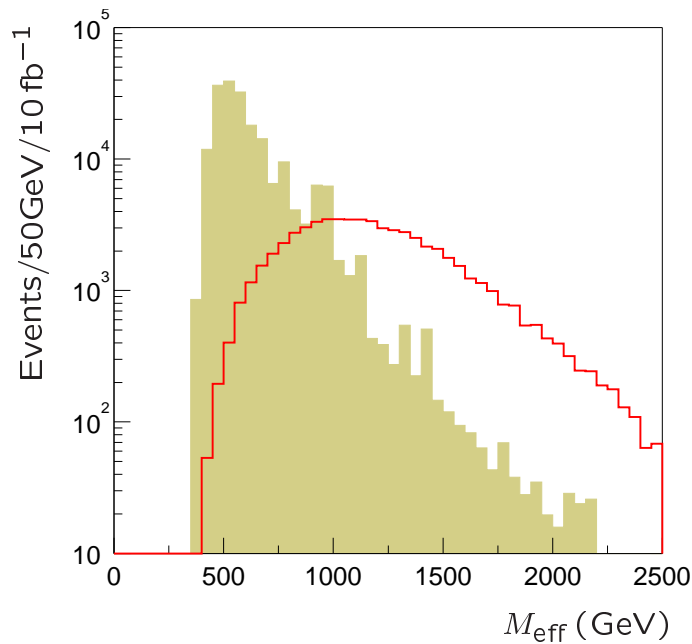}
\caption{$M_{\rm eff}$ distribution for a typical mSUGRA point and SM
backgrounds after cuts.} \label{fig:meff}
\end{figure*}

Due to the large production cross-sections, the SUSY particles
can be inclusively observed over the SM background
in the LHC data with very simple cuts.
The generic signatures are large missing transverse energy ($\not\!{E_T}$) 
and multiple hadronic jets and/or leptons. A typical example of
this signature is the distribution of the so-called effective mass,
\begin{eqnarray*}
M_{\rm{eff}} & = & \not\!{E_T} + \sum_{i=1}^{4} p_{T,i},
\end{eqnarray*}
i.e., the sum of the missing transverse energy and the transverse energy
of the four hardest jets. Its distribution is shown in Fig.~\ref{fig:meff}
together with the expected SM background for a mSUGRA model with
squark masses of approximately 700~GeV after requiring at least
four high-$P_T$ jets and significant missing transverse energy. 
The simulated data correspond
to an integrated luminosity of 10 fb$^{-1}$. It can be seen that for
large values of $M_{\rm{eff}}$ SUSY events can be selected with
negligible SM background.

\begin{figure*}[t]
\centering
\includegraphics[width=7cm,clip]{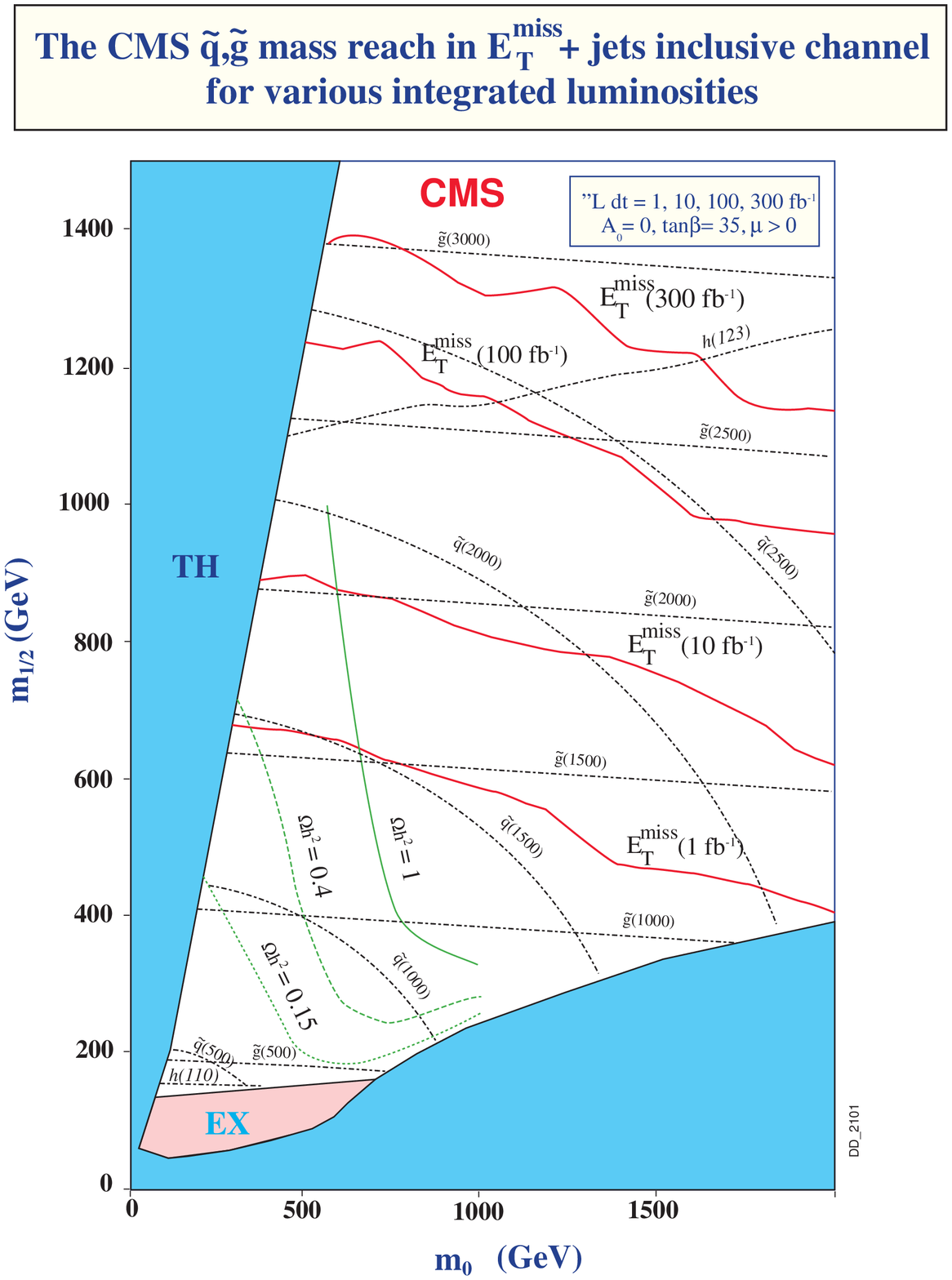}\quad
\includegraphics[width=7cm,clip]{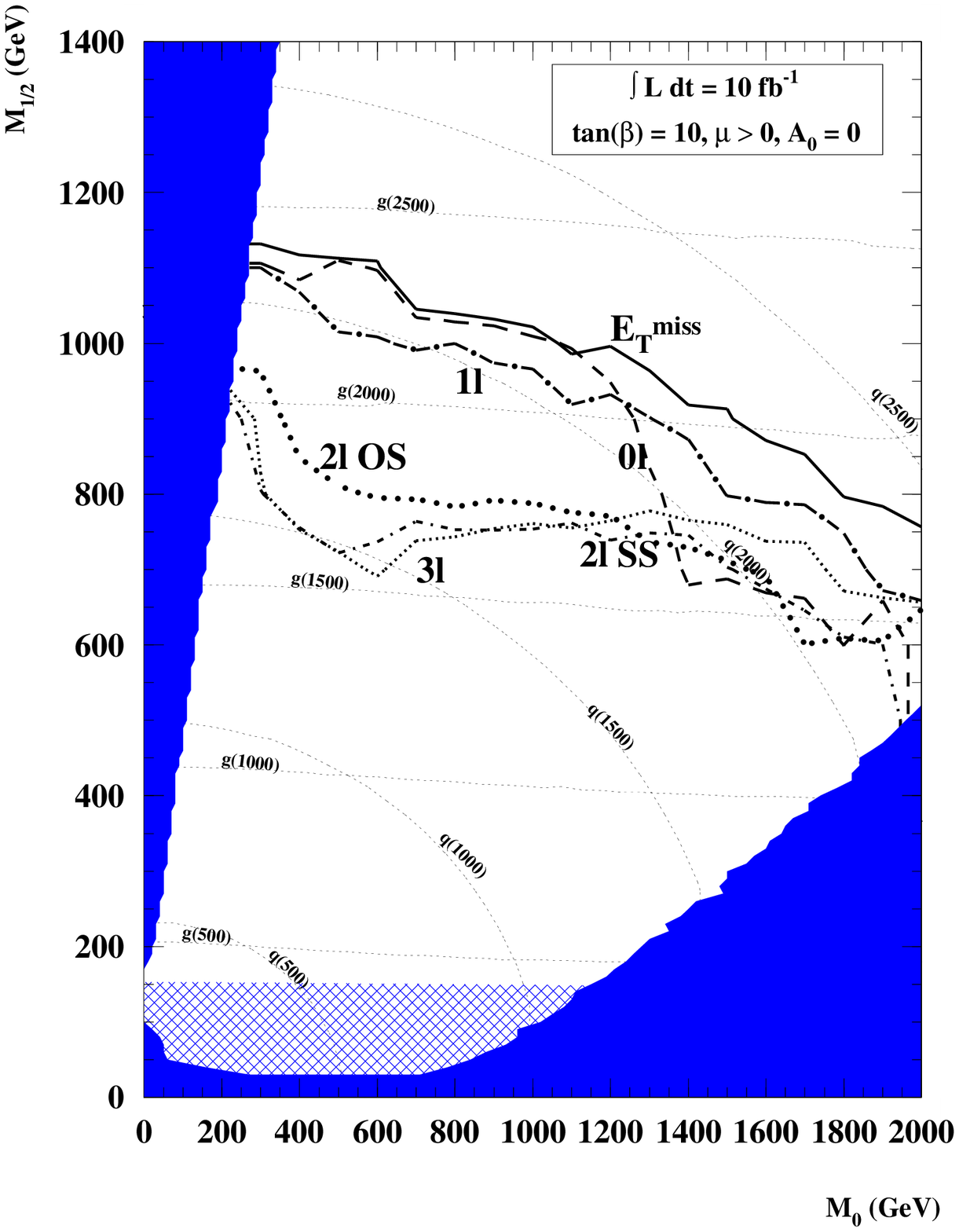}
\caption{Left: LHC reach in the $m_0-m_{1/2}$ plane for inclusive discovery 
of SUSY in the missing transverse
energy signature for the CMS experiment for an mSUGRA scenario
with $\tan\beta = 35$. Right: LHC reach for various inclusive SUSY 
signatures involving 0-3 leptons for 10 fb$^{-1}$.} \label{fig:susy_reach}
\end{figure*}

In Fig.~\ref{fig:susy_reach}(left) the 5$\sigma$ discovery reach in the
plane of the mSUGRA parameters $m_0$ and $m_{1/2}$ is shown for
integrated luminosities of 1,10,100,300 fb$^{-1}$ (red lines). 
The remaining parameters $A_0$ and $\tan\beta$ are fixed to 0 and 35,
respectively, and sgn($\mu$) is chosen positive. The discovery reach
depends only weakly on these parameters. Also shown are lines of
constant squark and gluino masses. For (1,10,300) fb$^{-1}$ the
mass reach for squarks and gluinos is approximately (1,2,2.5-3) TeV
thus covering a very large part of the mSUGRA parameter space. The
same applies qualitatively as well for a large part of the general
MSSM with neutralino LSP.

Under the assumption of GUT unification of the gaugino mass parameters
$M_1,M_2,M_3$ and sfermion mass parameters, the gluinos and squarks
are heavier than the color-neutral superpartners due to renormalization
group effects. Hence, the heavier charginos and neutralinos 
often appear in the decay chains of squarks. Their electro-weak decays
often give rise to high-$p_T$ leptons whose efficient detectability
provides an additional inclusive SUSY signature in many cases. 
In Fig.~\ref{fig:susy_reach}(right) the reach of the different lepton
signatures (1 lepton, 2 like-sign (SS), 2 opposite-sign (OS), 3
leptons) plus $\not\!{E_T}$ is shown for 10 fb$^{-1}$.

\subsection{Measurement of SUSY particle masses}

While inclusive detection of SUSY processes is quite straight-forward
at the LHC, the reconstruction of exclusive decay chains and the
reconstruction of superpartner masses is quite
involved due to various reasons: 1. the long decay chains of gluinos
and squarks lead to signatures with many jets and leptons with huge
combinatorics. 2. Due to the unknown longitudinal boost of the colliding
partons of the initial state no kinematic constraints from the beam
particles can be applied. 3. The event-by-event reconstruction of
superpartner invariant masses is not possible due to the undetected LSP's.
The reconstruction of superpartner masses has therefore to rely on the 
detection of kinematic endpoints in the invariant masses of the
detectable final state partons (jets and leptons) as well as on some
knowledge about the involved decay chains.

\begin{figure*}[htbp]
\centering
\includegraphics[width=9cm,clip]{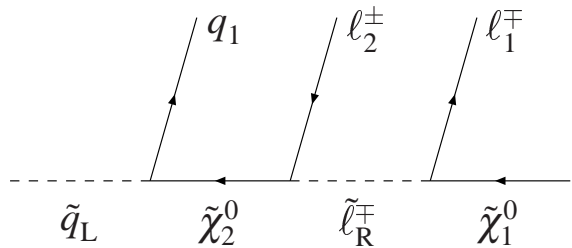}
\caption{Typical decay cascade of a left-chiral squark}
\label{fig:chain}
\end{figure*}

\subsubsection{The $\tilde q_L \to q\ell^+\ell^-\tchi^0_1$ decay chain}
\begin{figure*}[htbp]
\centering
\includegraphics[width=8cm,clip]{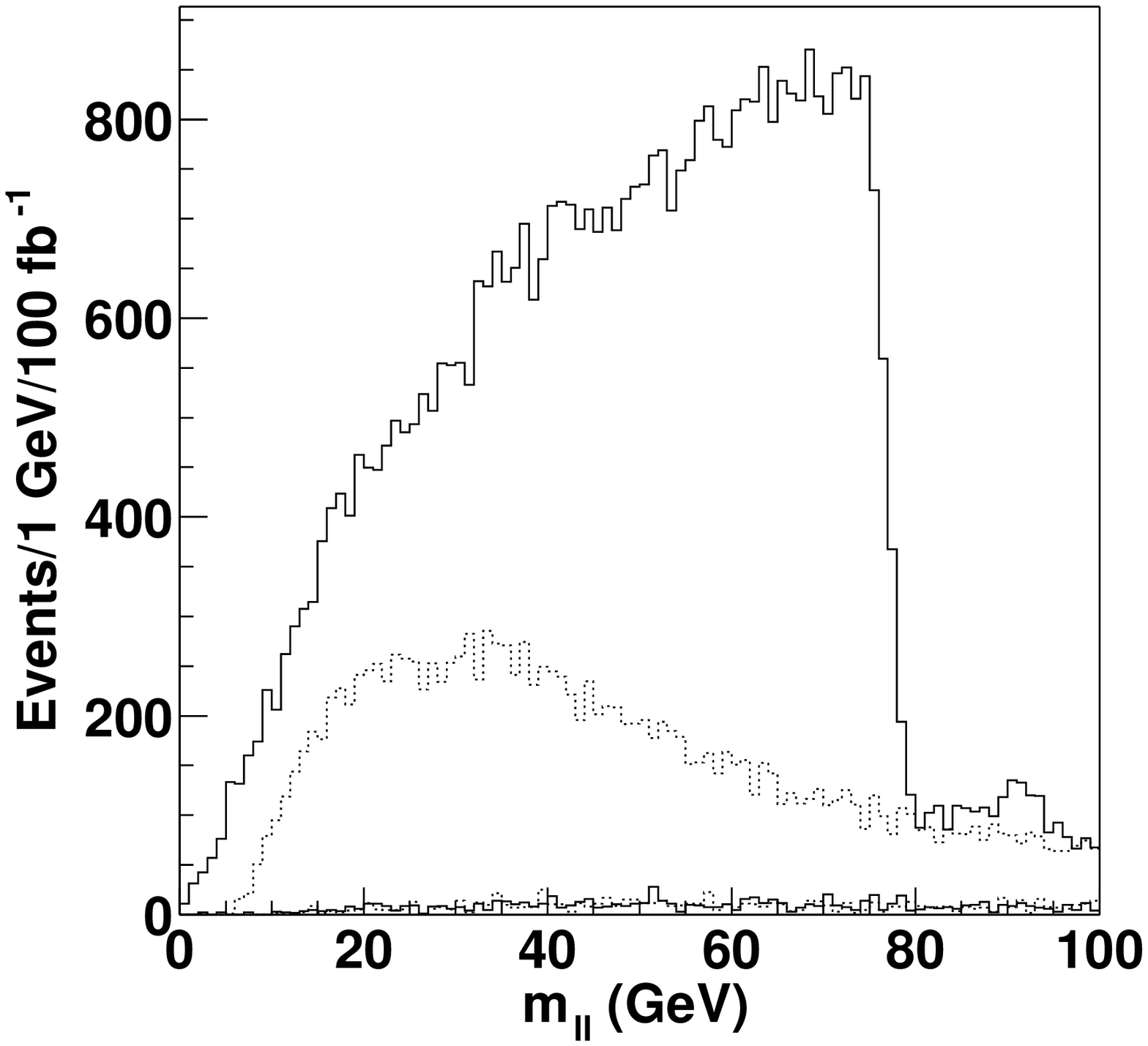}
\includegraphics[width=8cm,clip]{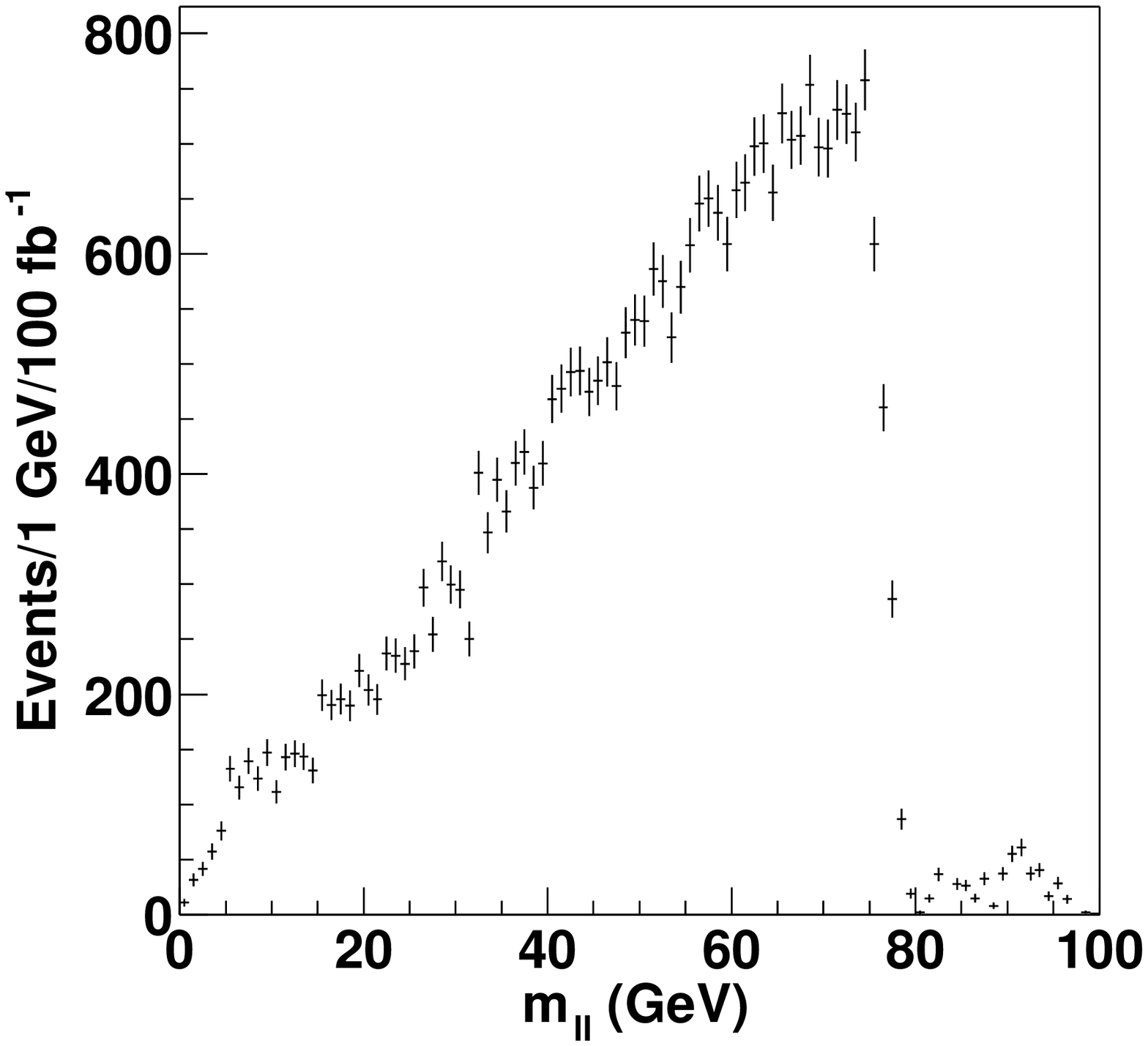}
\caption{Left: opposite sign same flavor (OS-SF) di-lepton 
mass spectrum. The upper curve is from 
$\tchi^0_2$ decays, the next curve is SUSY background, the lowest
curve is SM background. Right: same after subtraction of the
OS-OF rate which reduces the background from final states with
uncorrelated lepton flavour.}
\label{fig:mll}
\end{figure*}

A frequently
occurring and rather well reconstructible decay chain of a left-chiral squark
via the second-lightest neutralino $\tilde\chi^0_2$ and a right-chiral
slepton $\tilde\ell_R$ is shown in Fig.~\ref{fig:chain}. The invariant
mass distribution 
of the two opposite-sign same-flavor (OS-SF) leptons has a characteristic
triangular shape which exhibits a distinct
kinematic endpoint ('edge') which involves the unknown masses of
the three involved SUSY particles:
\begin{eqnarray*}
M_{\ell\ell}^{edge} =
\sqrt{(M_{\tchi_2^0}^2-M_{\tell}^2)(M_{\tell}^2-M_\lsp^2)}/M_\tell\, .
\end{eqnarray*}
In Fig.~\ref{fig:mll}(left) the (OS-SF) di-lepton signal from $\chi^0_2$
 decay is shown is shown together with the background from other SUSY
decays and the (negligible) SM background. The SUSY background
results mainly from wrongly combined and thus uncorrelated
leptons from independent neutralino or chargino decays. It 
can be efficiently determined from the rate of OS-OF di-leptons in the
data and then subtracted, as shown in the right part of Fig.~\ref{fig:mll}.
If the sleptons are heavier than $\tchi_2^0$, the three-body decay
$\tchi^0_2\to\ell^+\ell^-\tchi^0_1$ dominates which is different in
shape and has an endpoint at the neutralino mass difference,
$M_{\tchi^0_2} - M_{\tchi^0_1}$.

While from the di-lepton endpoint alone no absolute superpartner
masses can be extracted, further information can be obtained from the
squark decay chain shown in Fig.~\ref{fig:chain} from various
combinations of the leptons with a jet~\cite{lester,gjelsten}.
In particular, the following additional mass relations for kinematic edges can
be exploited:
\begin{figure*}[htbp]
\centering
\includegraphics[width=8cm,clip]{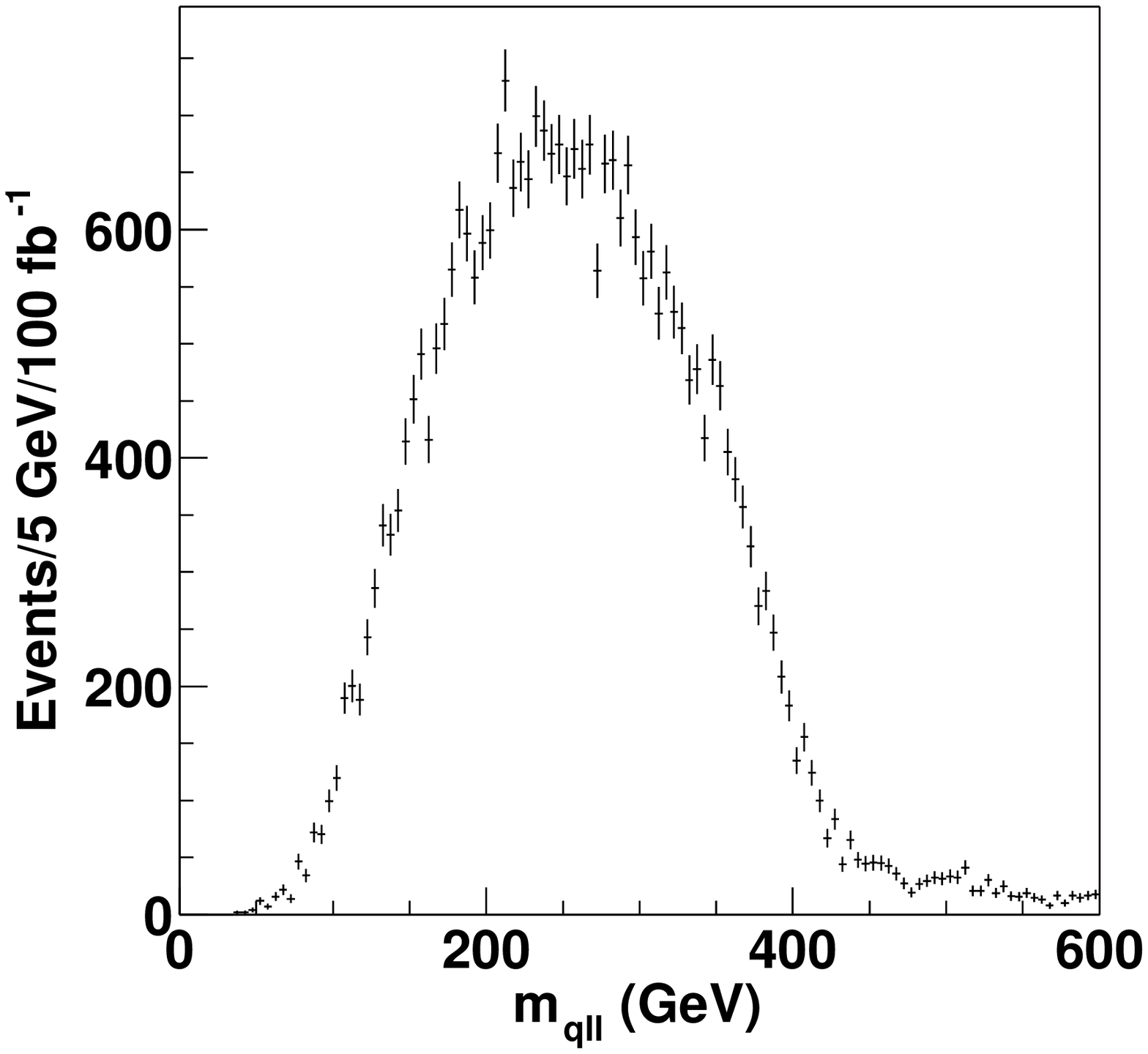}
\includegraphics[width=8cm,clip]{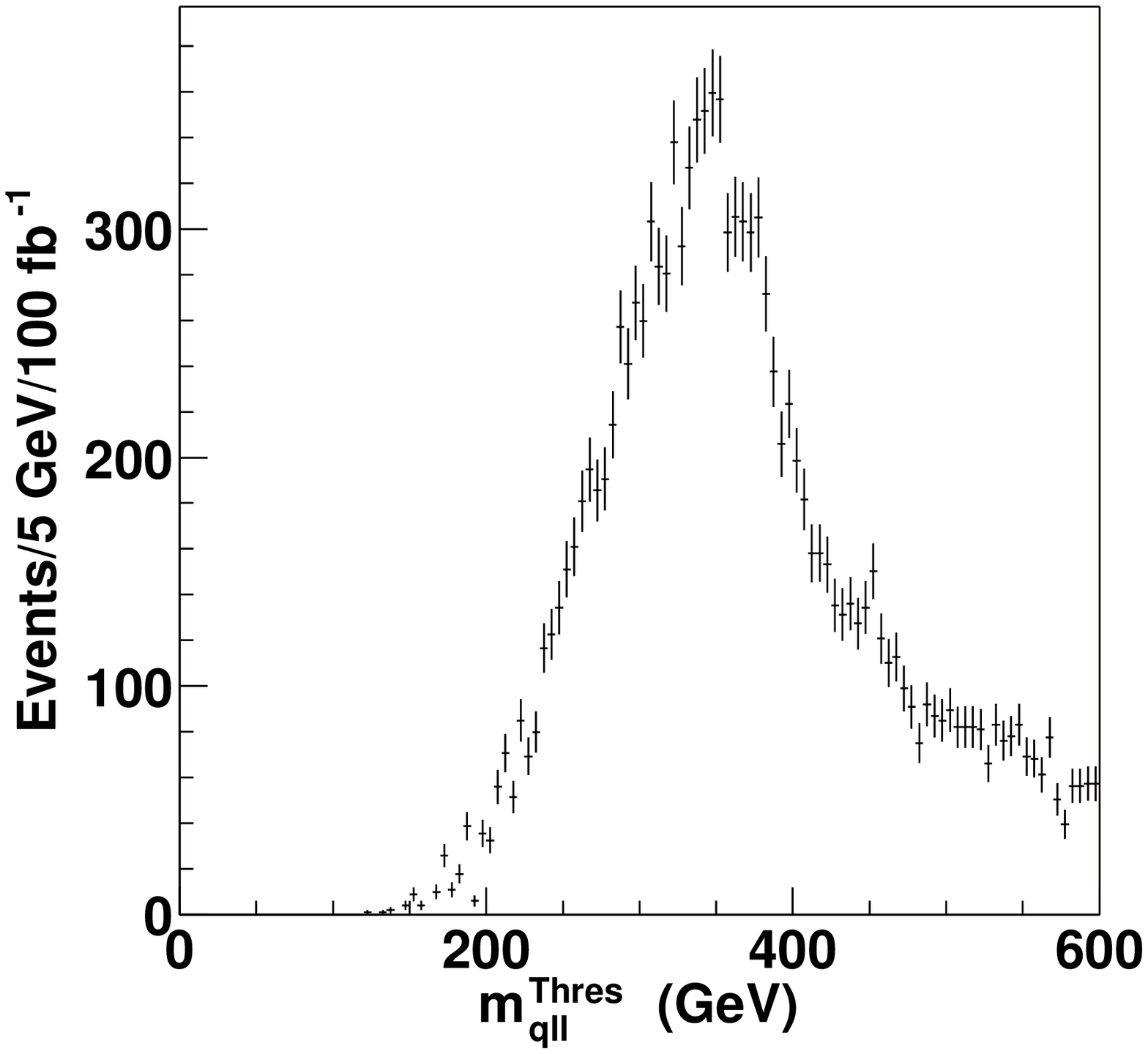}
\includegraphics[width=8cm,clip]{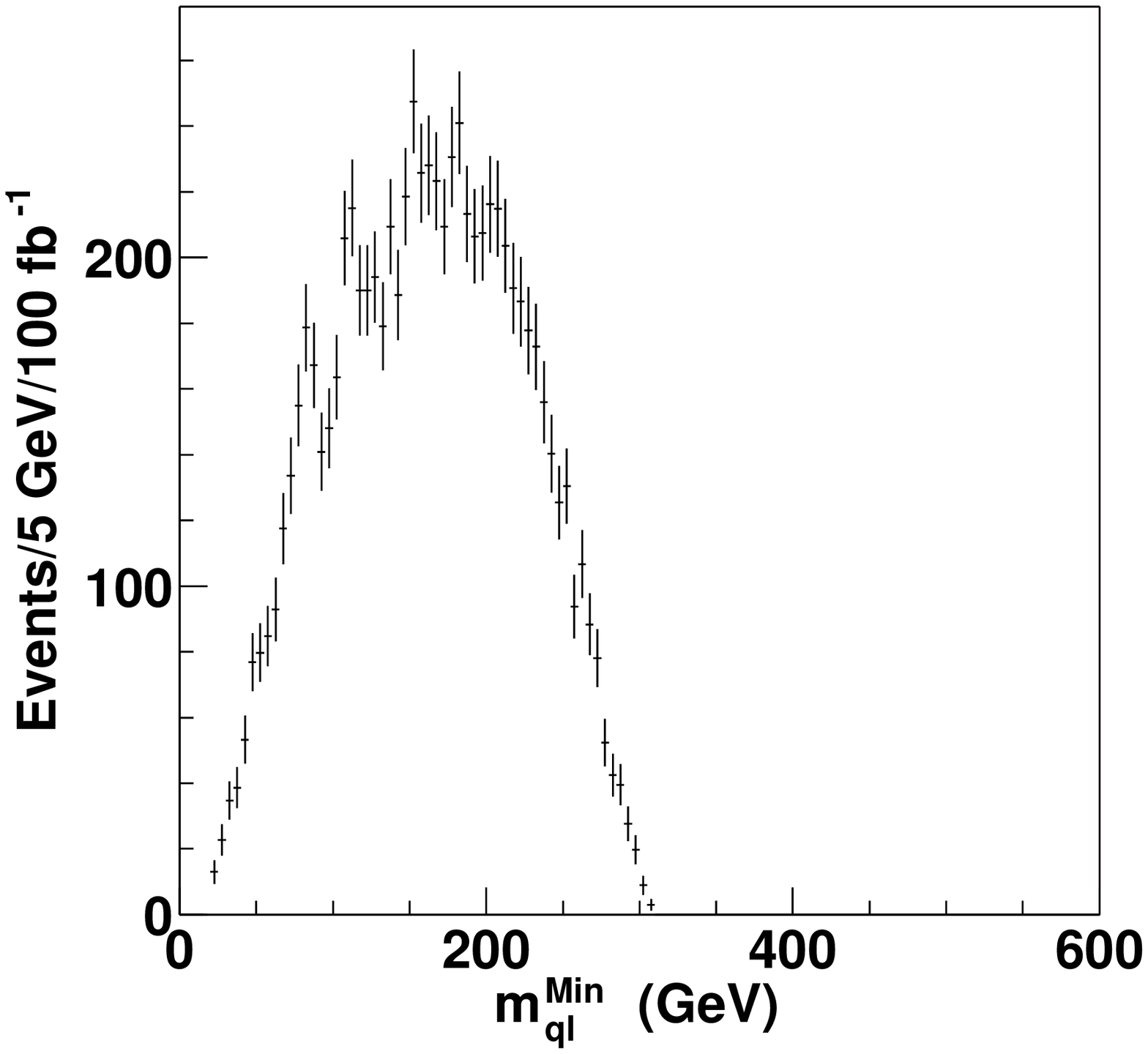}
\includegraphics[width=8cm,clip]{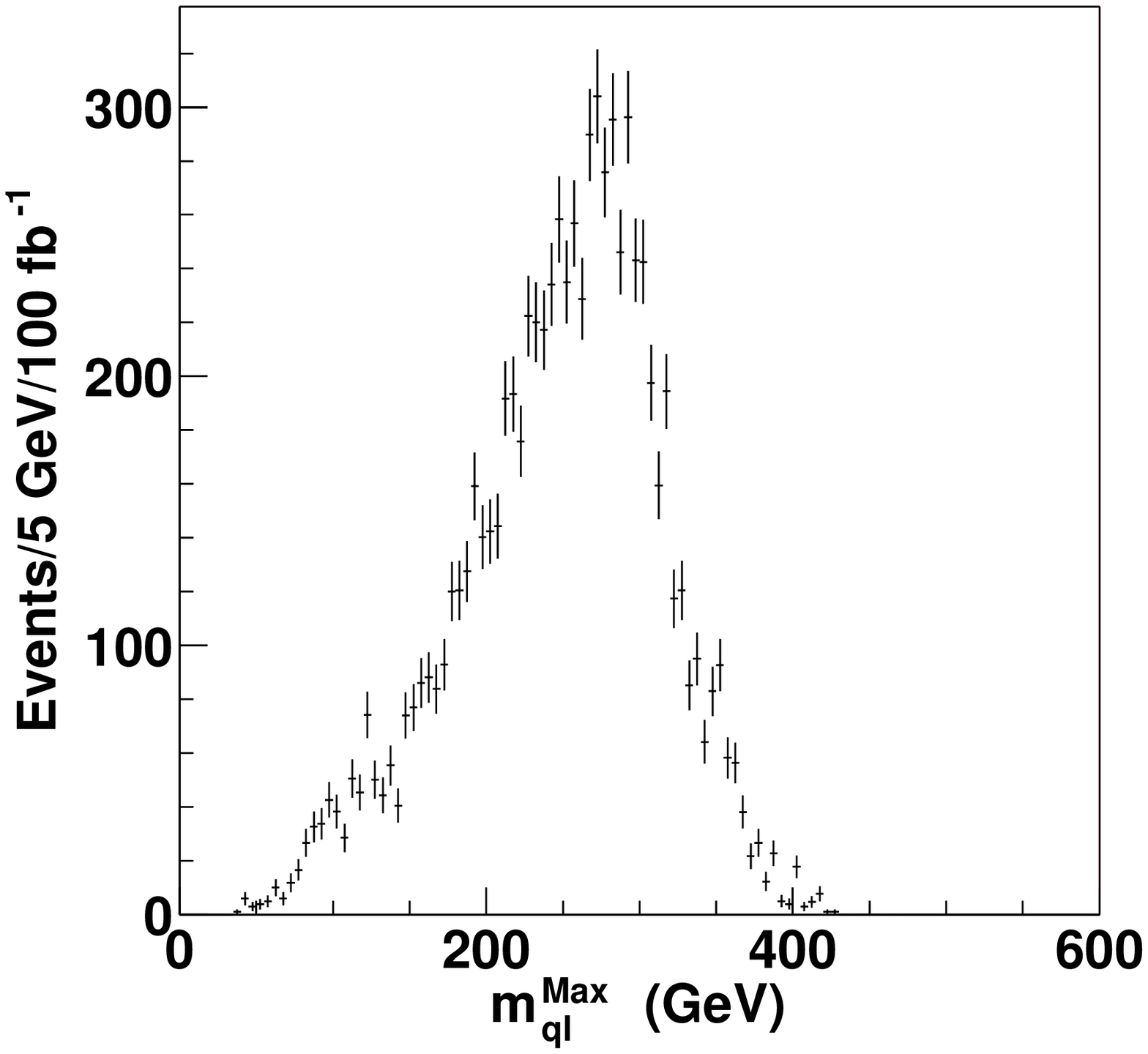}
\caption{Various mass spectra which exhibit kinematic endpoints
in squark decay cascades.}
\label{fig:edges}
\end{figure*}

\begin{eqnarray*}
\bigl(m_{q ll}^2\bigr)^{edge} 
&=& \frac{\bigl(\mq-\mT\bigr)\bigl(\mT-\mO\bigr)}{\mT}
\\
\bigl(m_{q l}^2\bigr)^{edge}_\min 
&=& \frac{\bigl(\mq-\mT\bigr)\bigl(\mT-\ml\bigr)}{\mT}
\\
\bigl(m_{q l}^2\bigr)^{edge}_\max 
&=& \frac{\bigl(\mq-\mT\bigr)\bigl(\ml-\mO\bigr)}{\ml}
\\
\bigl(m_{q ll}^2\bigr)^\threshold 
\nonumber
&=&[(\mq+\mT)(\mT-\ml)(\ml-\mO)\\
\nonumber
&&-(\mq-\mT)\sqrt{(\mT+\ml)^2(\ml+\mO)^2-16\mT\mlfour\mO}\\
&&+ 2\ml(\mq-\mT)(\mT-\mO)]/(4\ml\mT)
\end{eqnarray*}
The labels ``min'' and ``max'' refer to the distribution 
constructed from the smaller and the larger of the two $q\ell$ masses.
Furthermore ``thres'' refers to the threshold in the subset
of the $m_{qll}$ distribution for which the angle between the
two lepton momenta (in the slepton rest frame) exceeds $\pi/2$,
which corresponds to $m_{ll}^\edge/\sqrt{2}<m_{ll}<m_{ll}^\edge$.
The corresponding mass distributions are shown in Fig.~\ref{fig:edges}.
The position of the di-lepton edge can be measured to a statistical 
precision of better than 100 MeV and the edges involving jets can be
measured to few GeV precision with 100 fb$^{-1}$ of data. In the case
of the latter a systematic uncertainty from jet energy scale of 
approximately 1\% has to be accounted for as well.
Since the number of measurable edges in this scenario is larger than
the number of involved superpartner masses, the absolute masses can be
extracted from a simultaneous fit. The achievable precisions are
listed in the first column of Table~\ref{tab:susymasseslhclc}.

\subsubsection{Gluino and third generation squarks}

\begin{figure*}[htbp]
\centering
\includegraphics[width=5.3cm,clip]{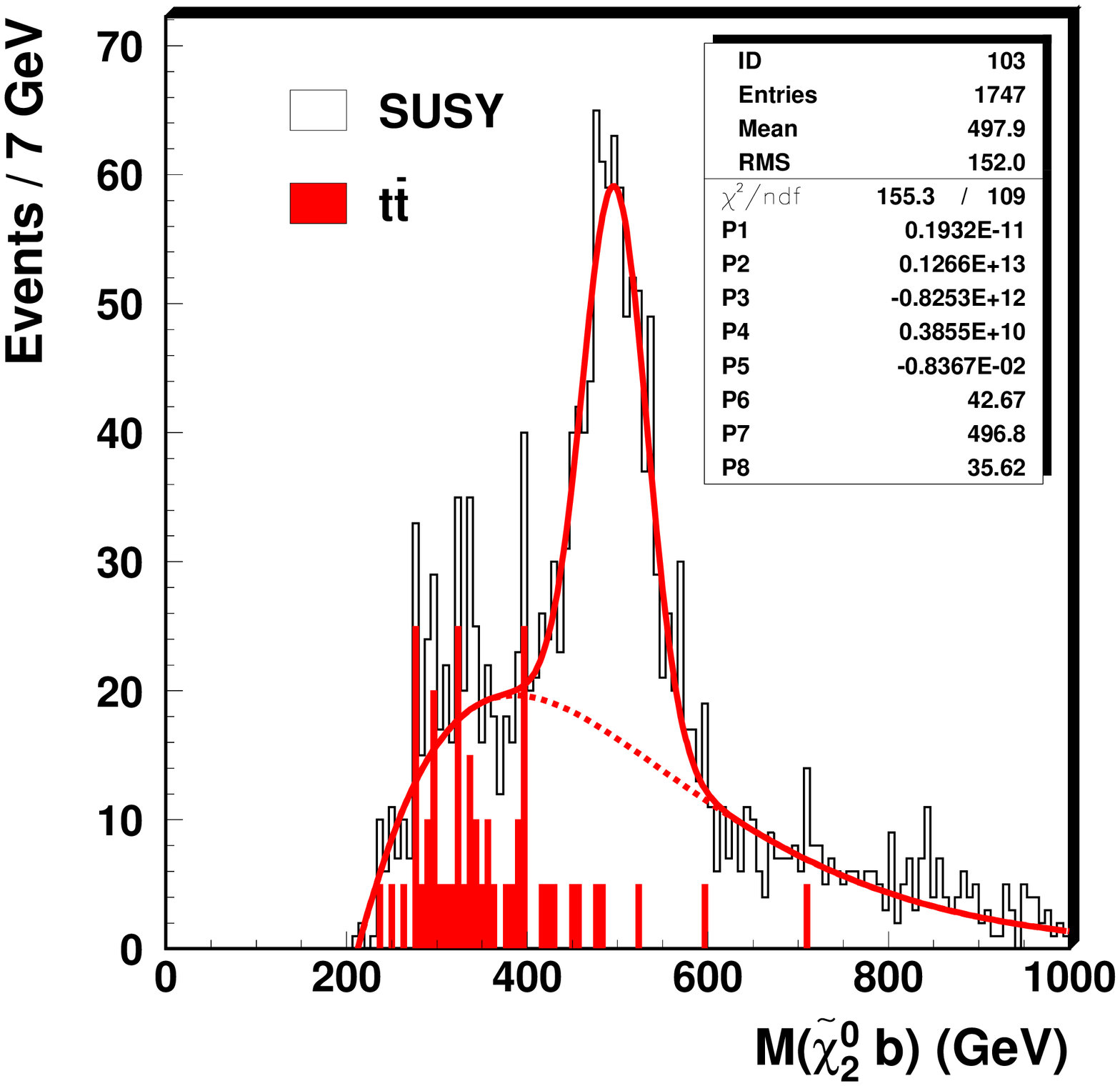}\quad
\includegraphics[width=5.3cm,clip]{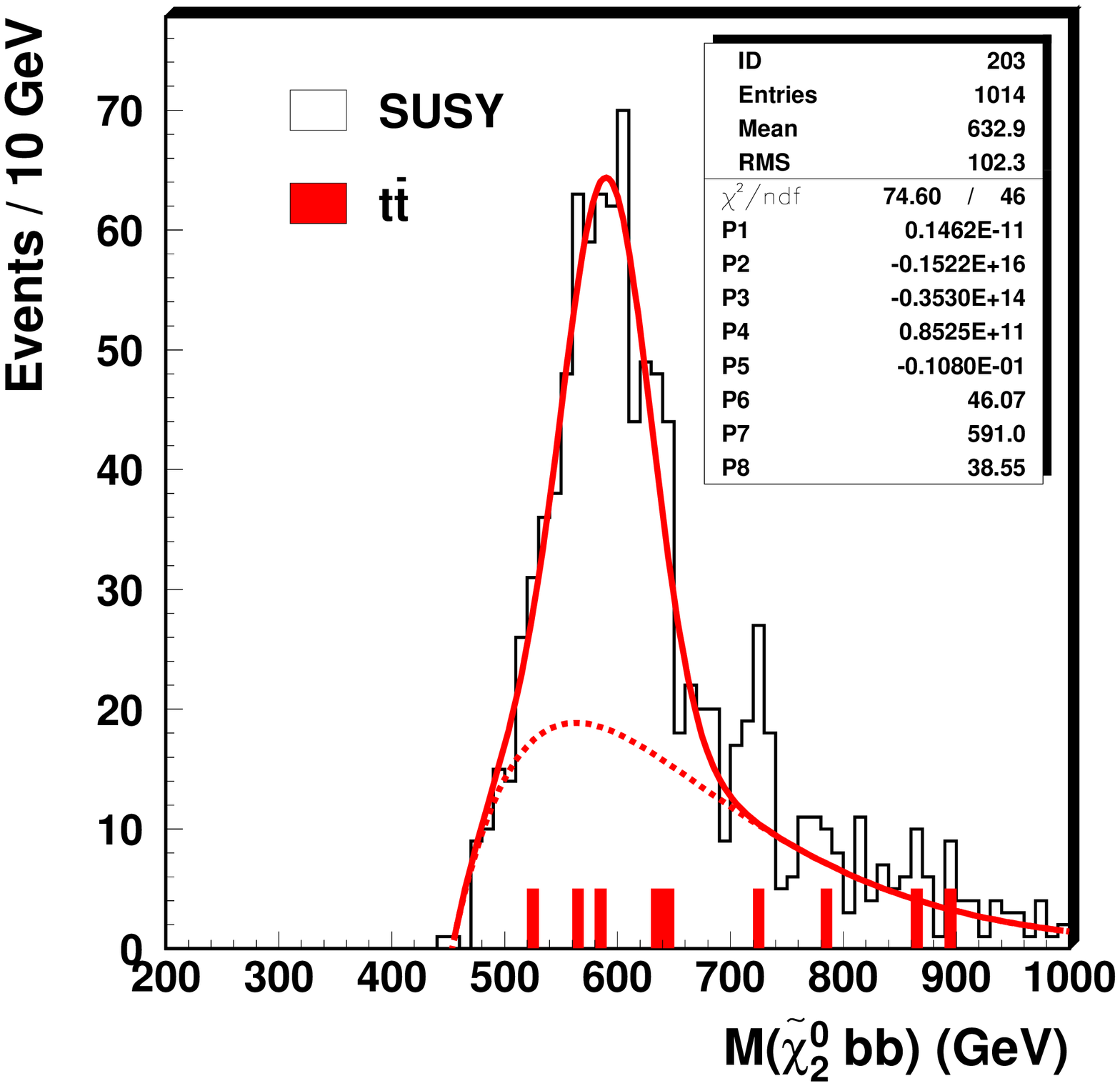}\quad
\includegraphics[width=5.3cm,clip]{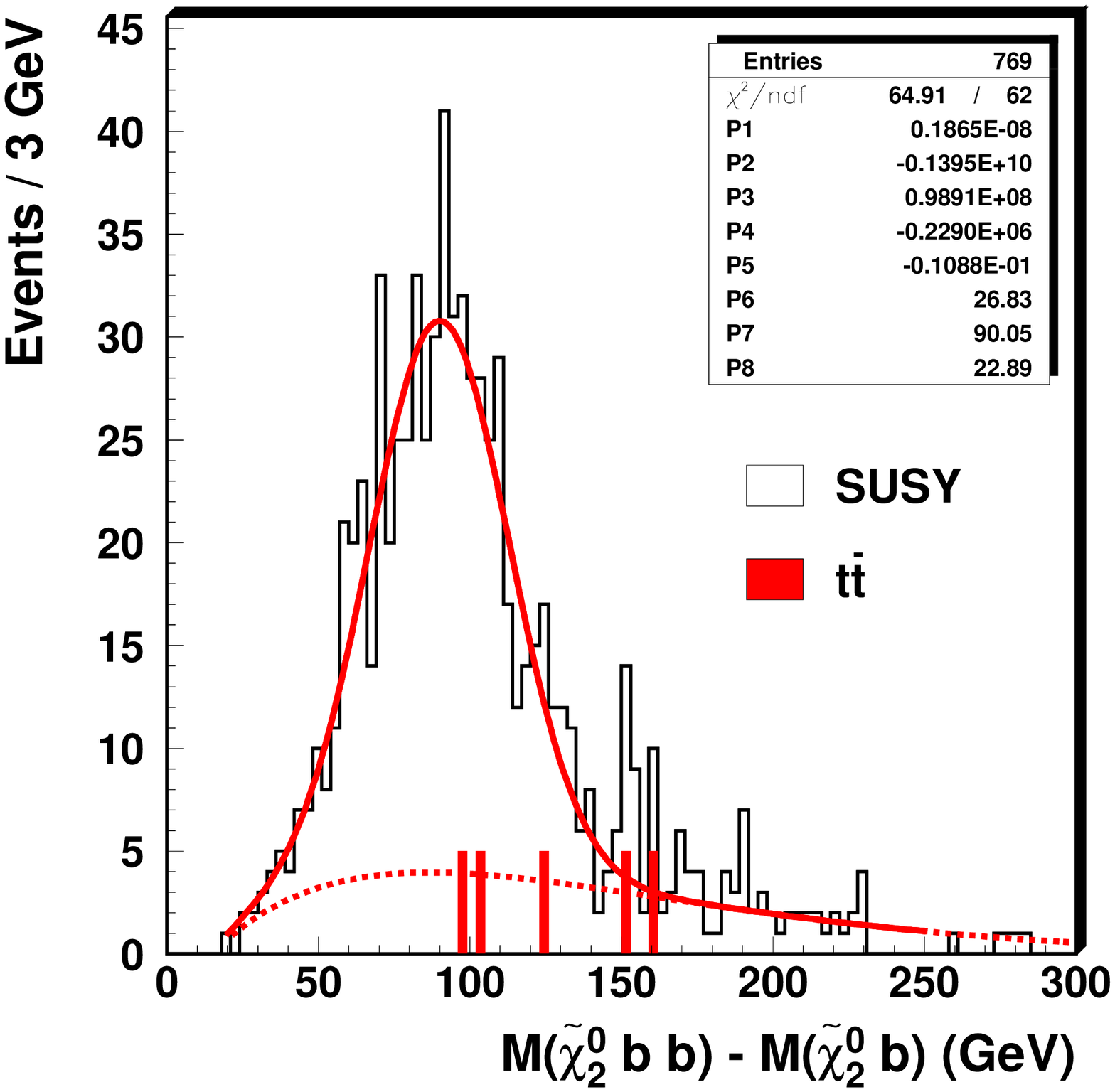}
\caption{Reconstructed mass distribution for sbottom (left),
  gluino (middle), and for the gluino-sbottom mass difference (right)
(from~\cite{glusbot}).}
\label{fig:sbotglu}
\end{figure*}

Knowing the masses of $\tchi^0_1, \tchi^0_2, \tl_R$, and $\tq_L$,
gluinos can be reconstructed by adding another quark jet to the 
$q\ell\ell$ system. In particular when the decay proceeds through a
$\tilde b$ squark, the tagging of two b-jets reduces combinatorial
background. Knowing the mass of the LSP (from the joint fit to the
kinematic edges described above), the momentum of the $\tchi^0_2$ can
be approximated by
\begin{eqnarray*}
\vec{p}(\tchi^0_2)=\left( 1-\frac{m(\lsp)}{m(\ell\ell)}\right)\vec{p}_{\ell\ell}\end{eqnarray*}
if the $\tchi^0_1$ carries negligible momentum in the $\tchi^0_2$ rest
frame, which in the SPS1a scenario is the case for events close to
kinematic endpoint of the di-lepton mass spectrum.
Knowing the $\tchi^0_2$ momentum and mass, the sbottom mass can be
subsequently reconstructed as the $b\tchi^0_2$ invariant
mass and the gluino mass as the $b\bar{b}\tchi^0_2$ invariant mass.
The reconstructed mass distributions in CMS for sbottom and gluino are shown
for 300~fb$^{-1}$ are shown in Fig.~\ref{fig:sbotglu} (left, middle).
The mass difference 
between gluino and sbottom can be reconstructed without assumptions on
the $\tchi^0_1$ mass (Fig.~\ref{fig:sbotglu}, right)~\cite{glusbot}.

The reconstruction of $\tilde t$ squarks is more challenging. Initial
studies are available~\cite{stoplhc}.

\subsubsection{ Mass relation method}

An alternative approach to mass reconstruction in the presence of long
decay chains at the LHC, the {\it mass relation method}
has recently been proposed~\cite{massrelation}. If one considers 
e.g.~the decay chain $\tilde g \to \tilde b b_2 \to \tchi^0_2 b_1 b_2 \to
\tilde \ell b_1 b_2 \ell_2 \to \tchi^0_1 b_1 b_2 \ell_1 \ell_2$, the
five superpartner mass $m_{\tchi^0_1}, m_{\tilde \ell}, m_{\tchi^0_2},
m_{\tilde b}, m_{\tilde g}$ can be calculated from the 5 4-momenta of the
final state particles all of which except for $p_{\tchi^0_1}$ are measured.
Thus, the set of superpartner masses compatible with a single observed
event corresponds to a 4-dimensional hypersurface in the 5-dimensional 
mass space. Since the exact location of the hypersurface is different for each
event, the ensemble of hyper-surfaces from all events will have intersections
at the true values of the five unknown masses. This method has been applied
in a simplified version to the reconstruction of sbottom and gluino mass
under the assumption that the other masses are known, in which case the
hypersurface reduces to a line in ($m_{\tilde g}, m_{\tilde b}$) space,
i.e. for each pair of events a solution for ($m_{\tilde g}, m_{\tilde b}$)
is obtained up to a two-fold ambiguity.
The advantages of the method are that it is based on exact kinematics without
any approximation and that all events, not only the ones close to the kinematic
endpoint, can be used. Furthermore, mass peaks are reconstructed rather than
kinematic edges. 

\subsubsection{Sleptons}

If sleptons are sufficiently light, they are produced at a decent rate in
pp collisions through Drell-Yan pair production. For the SPS1a benchmark,
the masses of the left- and righthanded selectrons and smuons are 143~GeV
and 202~GeV, respectively.
The total cross-section for smuon/selectron pair production is 91~fb. 
The signature are two opposite sign same flavor leptons, missing 
transverse energy and no jets. After subtraction of the opposite flavor
background only a few events remain for 100 fb$^{-1}$. However, from a
transverse mass estimator, the slepton mass can be estimated to a few GeV
precision under favorable circumstances~\cite{lytken}.

Staus may frequently occur in the decays of $\tchi^0_2$ if the $\tilde\tau_1$
is lighter than $\tchi^0_2$, which is the case in SPS1a and in generally
in mSUGRA models with significant $\tilde\tau$ mixing at large $\tan\beta$.
Hadronic $\tau$ decays can be tagged typically at an efficiency of 50\% for
a QCD jet rejection factor of 100 at low luminosity. Selecting di-$\tau$ events 
with $\not\!{E_T}$ and large $M_{\rm eff}$ and again subtracting the same sign
contribution, the invariant mass of the $\tau\tau$ decay products carries
information about the $\tilde\tau$ mass in its endpoint. A mass estimate 
with a couple of GeV precision seems feasible but further study is needed.

\subsubsection{Charginos}

Recently a method has been proposed to observe the lighter chargino, 
$\tchi^\pm_1$, frequently appearing in the decay chain of a left squark,
$\tilde q_L \to \tchi^\pm_1 q$ via its decay $\tchi^\pm \to W^\pm \tchi^0_1
\to q\bar{q}^\prime \tchi^0_1$. The methods uses di-lepton events where
the two leptons arise from the decay of the initial squark decaying
into $\tchi^0_2$ as described above. Assuming known masses for $\tchi^0_1,
\tchi^0_2, \tilde \ell_R,$ and $\tilde q_L$, the momentum of the $\tchi^0_1$
from the $\tchi^0_2$ can be reconstructed up to a two-fold ambiguity. 
After identification of hadronic $W$ bosons, the mass of the chargino can be 
reconstructed from the $W$ momentum, the reconstructed opposite-side 
$\tchi^0_1$ and the total $\not\!{E_T}$ up to another two-fold ambiguity. 
After background subtraction of events in the side-bands of the reconstructed 
$W$ mass, a peak becomes visible in the reconstructed $\tchi^\pm_1$ mass
distribution. For a mSUGRA scenario with $m_0 = 100$ GeV, $m_{1/2} = 300$ GeV,
$A_0 = -300$ GeV, $\tan\beta = 6$, and sgn$(\mu) = +$, a 3$\sigma$ excess
is achievable for 100 fb$^{-1}$ and a mass estimation with 
approximately 10\% error seems feasible~\cite{chargino}.

\subsubsection{Heavy gauginos}

The left squarks predominantly decay via 
the gaugino-like neutralinos and charginos, i.e.~usually $\tchi^0_2$ and
$\tchi^\pm_1$ in mSUGRA models. While $\tchi^0_3$ often is pure Higgsino,
the heaviest neutralino $\tchi^0_4$ and the heavy chargino $\tchi^\pm_2$
have some gaugino admixture leading to a production of $\tchi^0_4$ and
$\tchi^\pm_2$ in the $\tilde q_L$ decays with a branching ratio of a few 
percent. Four different decay chains via left and right sleptons (for the
neutralino) and via a sneutrino (for the chargino) lead to di-lepton
signatures of correlated flavor and charge with di-lepton 
masses significantly larger than for the $\tchi^0_2$ decay described above. 
While it seems very hard to disentangle the four corresponding kinematic 
end-points, the observation of heavy gaugino production is possible with
a dedicated analysis in a part of the mSUGRA parameter space, if $m_0$ is
not too large. In particular, the highest mass end-point can be
measured to a precision 
of approximately 4~GeV for the SPS1a scenario with 100 fb$^{-1}$ of
data. Its unambiguous identification
needs additional information, however~\cite{heavygauginos}.

\subsection{Measurement of the $\tchi^0_2$ Spin}
\label{sec:lhcspin}

\begin{figure*}[htbp]
\centering
\includegraphics[width=8cm,clip]{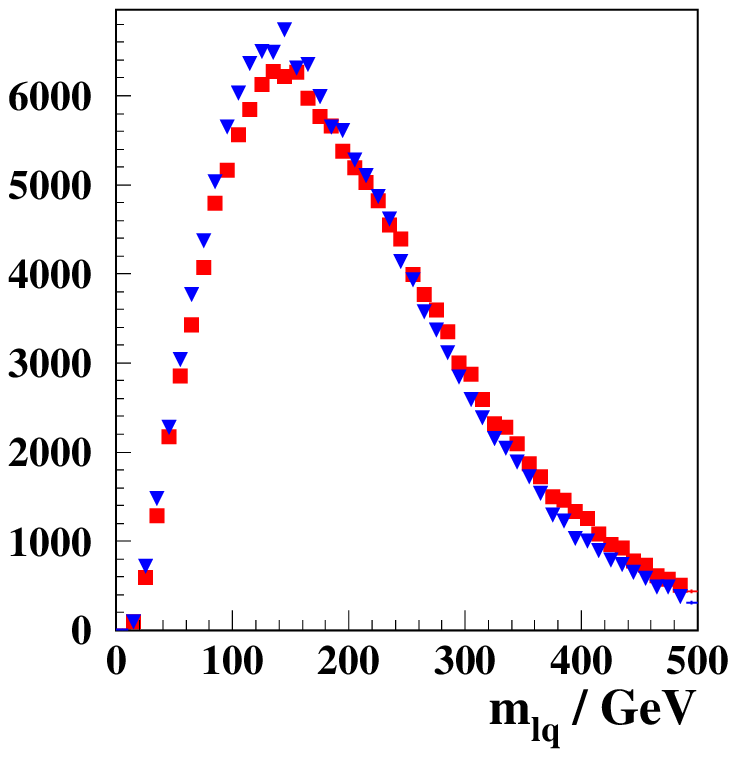}
\includegraphics[width=8cm,clip]{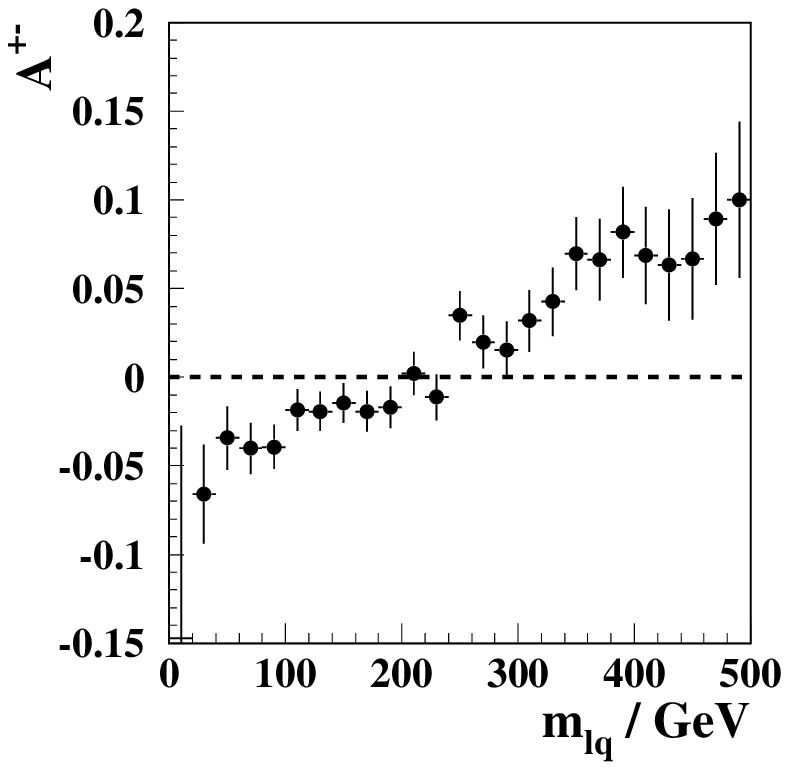}
\caption{Left: Distribution of the charged lepton - jet invariant mass
distribution for positive and negative leptons after detector simulation. 
Right: Observable charge asymmetry in the charged-lepton mass spectrum
after detector simulation for 150 fb$^{-1}$ (from~\cite{lhcspin}).}
\label{fig:lhcspin}
\end{figure*}

The reconstruction of the superpartner spins is together with the determination
of the their gauge quantum numbers the crucial test of SUSY. 
Spin information is carried by the scattering angle distribution of the
primary pair of superpartners in their rest-frame. In hadron collisions, 
it is however hard to determine this frame due to the unknown longitudinal
boost of the partonic initial state and due to the unobserved LSP's in the
final state. It is however possible to exploit angular distributions of
the superpartner decay products in decay chains to obtain spin 
information. A particular example for a measurement of the $\tchi^0_2$
spin has been worked out, again exploiting the 
$\tilde q_L \to q\ell^+\ell^-\tchi^0_1$ cascade (see Fig.~\ref{fig:chain}).
Since the left squark always decays into a left-handed quark, the $\tchi^0_2$
becomes polarized. Since in its decay the $\tchi^0_2$ emits a scalar right 
slepton, the corresponding so-called {\it near} lepton will carry the 
polarization information of the $\tchi^0_2$. Therefore one expects a charge
asymmetry in invariant mass distribution of the quark and the near lepton,
$m(\ell_{near} q)$. For $\ell^+ q$ and $\ell^- \bar{q}$, 
the tree-level differential form the $m(\ell_{near} q)$ spectrum 
is $dP/dm \propto 4m^3$ while for
$\ell^- q$ and $\ell^+ \bar{q}$ it is $dP/dm \propto 4m(1-m^2)$.
These tree level distributions are not experimentally observable since
anti-quarks cannot be distinguished from quarks and the near lepton 
cannot be distinguished from the far lepton. If the original (anti)-squarks
are (at least partially) produced by the $qg\to\tilde q \tilde g$ and
$\bar{q}g\to\tilde{\bar{q}}\tilde g$, the p.d.f. asymmetry due to the valence
quarks in the protons can be exploited and more squarks than anti-squarks 
will be produced. The charge asymmetry from the far lepton is expected to
be very small. Thus, forming both the invariant mass with the near and the far
lepton dilutes the charge asymmetry but does not remove it. 
In Fig.~\ref{fig:lhcspin}(left) 
the reconstructed lepton-jet invariant mass distribution
is shown for positive (squares) and negative (triangles) lepton charge. 
On the right, the resulting charge asymmetry is shown for 150 fb$^{-1}$~\cite{lhcspin}.

\section{SUSY AT THE LINEAR COLLIDER}
\label{sec:lc}

\subsection{Experimental environment}

The International Linear Collider (ILC) 
is a projected electron positron collider
at 500-1000 GeV center-of-mass energy with a luminosity of 
several $10^{34}{\rm cm}^{-2}{\rm s}^{-1}$~\cite{TESLA,NLCreport,acfareport}.
The energy will be tunable from the Z-pole
up to the highest energy. Both beams can be polarized ($e^-$: 90\%, 
$e^+$: 50-60\%). The advantage of $e^+e^-$ collisions are the
known, electro-weakly interacting initial state, the low level of 
instrumental and physics backgrounds and as a consequence the
comparably low event
rate which allows to record all collisions without any trigger
requirements. Due to the high luminosity, unlike 
previous $e^+e^-$ machines, the ILC will have
significant beam-beam interactions. These lead to a production of
approximately 6$\times 10^{10}$ low-energetic photons per
bunch-crossing at 500~GeV. 
About 10\% of the events will have center-of-mass energies below 
95\% of the nominal energy. Therefore the beam-strahlung spectrum has
to be monitored continously with data (e.g.~acollinearity of Bhabha
events) and corrected for. Backgrounds from collisions of
beam-strahlung photons have been studied and were found to be small
except for the very forward region for which dedicated highly-granular
calorimeters have to be build.

\subsection{Mass measurements}

At the ILC the masses of the color-neutral superpartners can be
measured in two different ways. First in continuum production,
kinematic end-points and energy spectra 
can be used to extract simultaneously the
involved masses. Second, the measurement of the shape of the
production cross-section for various processes near threshold 
allows for a very precise extraction of the sum of the produced
superpartner masses.

\subsubsection{Sleptons}

Sleptons are pair-produced in the reactions
\begin{eqnarray*}
   e^+e^- & \to & \tilde\ell^+_i\tilde\ell^-_j, \, 
  \tilde\nu_\ell\,\bar{\tilde\nu_\ell}  \hfill
        \qquad\qquad\qquad \ell = e,\, \mu,\, \tau \ {\rm \ and \ } \
            [i,j = L,R \ {\rm \ or \ } \ 1,2]
\end{eqnarray*}
via $s$-channel $\gamma/Z$ exchange and $t$-channel $\tilde\chi$ 
exchange for the first generation. 

\begin{figure*}[htbp]
\centering
\includegraphics[height=5cm]{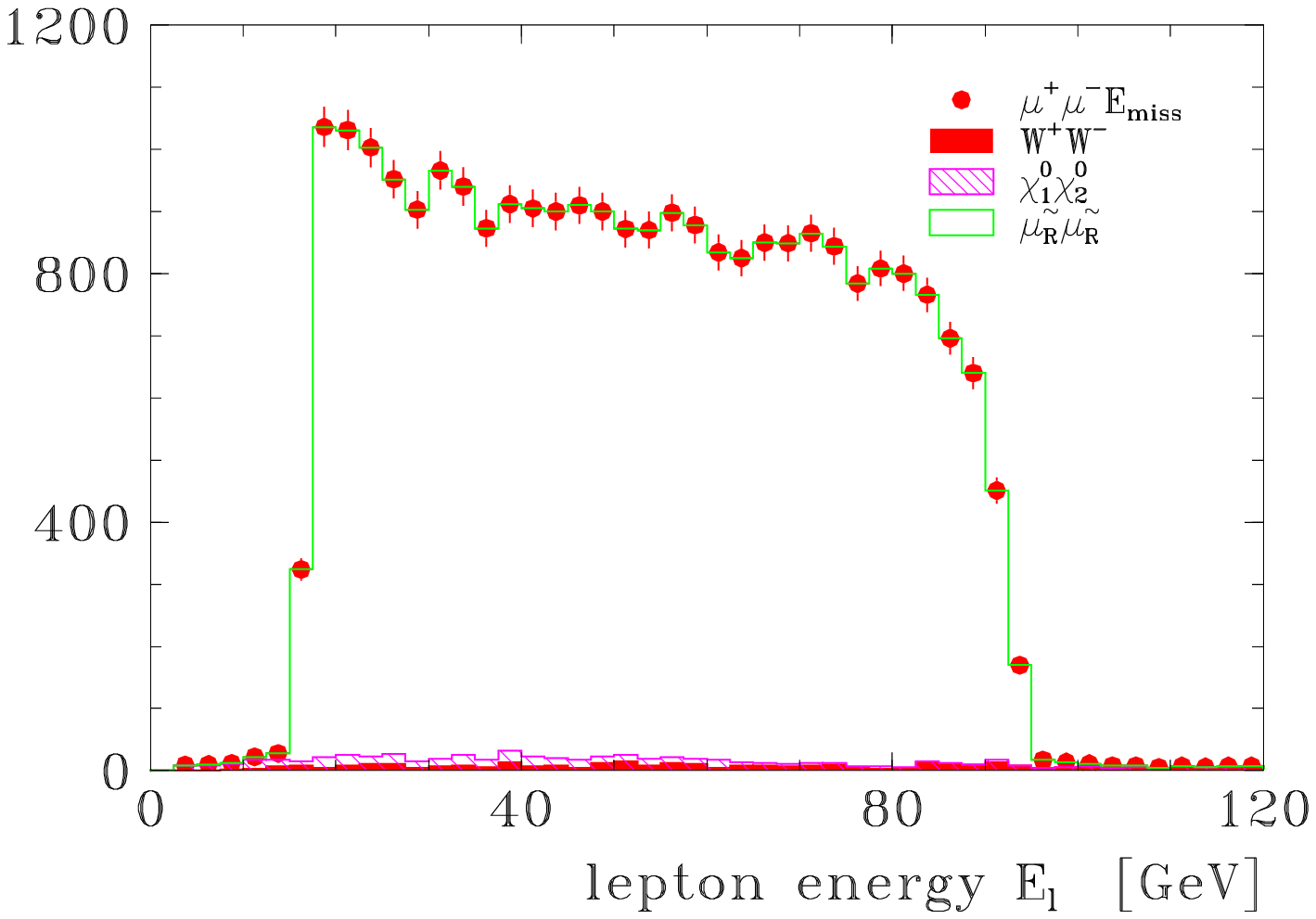}
\includegraphics[bbllx=85pt,bblly=230pt,bburx=300pt,bbury=390pt,clip,height=5cm]
{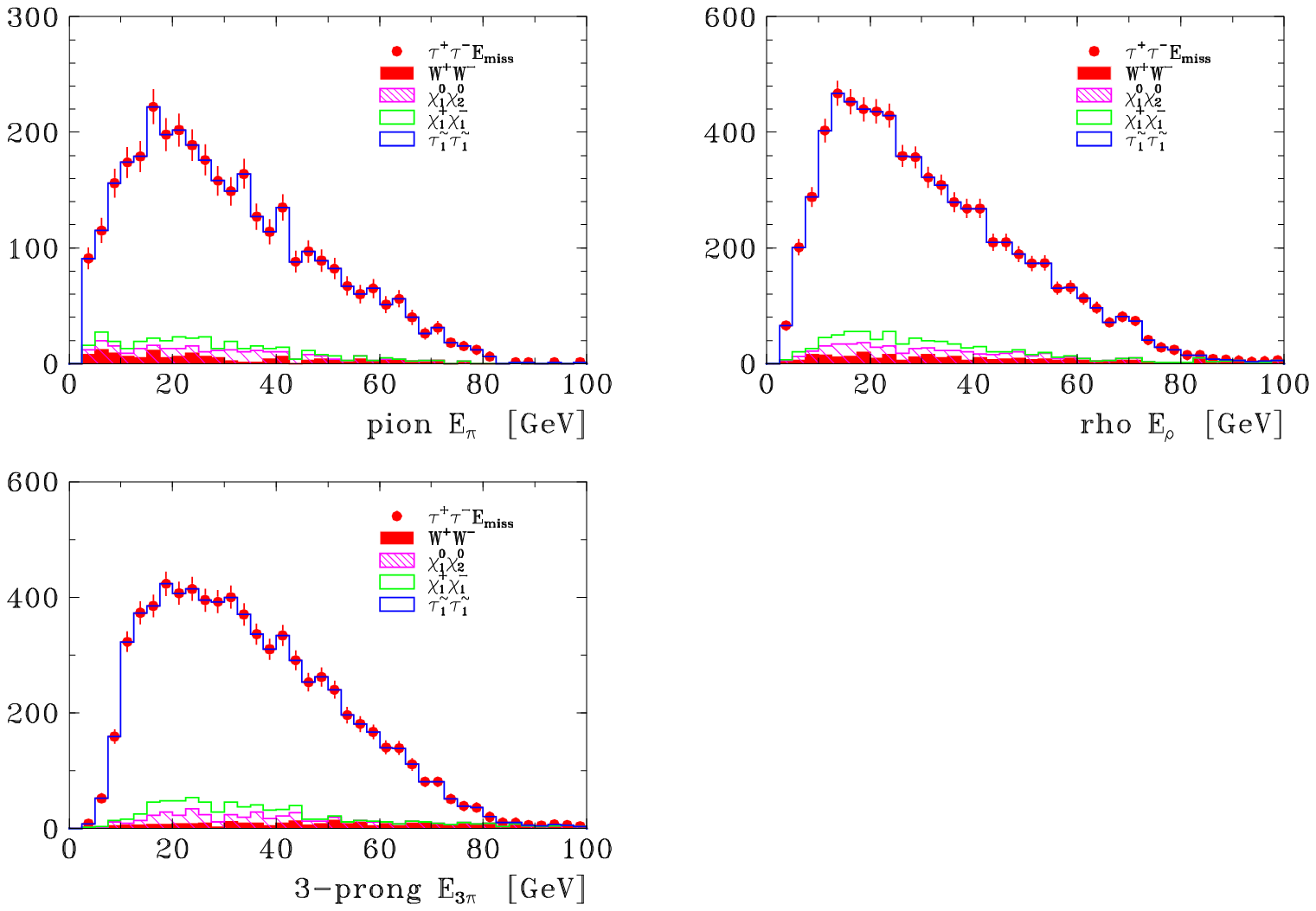}
\caption{Left: Muon energy spectrum from the process 
$e^+_L e^-_R\to\tilde{\tau}_1\tilde{\tau}_1\to\mu^+\tilde{\chi}^0_1
\mu^-\tilde{\chi}^0_1$
Right: Hadron energy spectrum of the decay 
$\tau\to 3\pi \nu_{\tau}$
from the process $e^+_L e^-_R\to\tilde{\tau}_1\tilde{\tau}_1$}
\label{fig:boxes}
\end{figure*}

As an example, in Fig.~\ref{fig:boxes}(left) the measurable energy spectrum 
of the muons from the process $e^+_Le^-_R\to\tilde\mu_R^+\tilde\mu_R^-
\to \mu^+\tchi^0_1\mu^-\chi^0_1$ is shown~\cite{sleptons}. 
Events can be selected with
negligible SM background. In particular background from W-pair
production can be efficiently suppressed by choosing right-handed
electrons in the initial state. SUSY backgrounds in this final state
are generally small and can be suppressed in part by topological cuts.
Due to the scalar nature of the smuons, the energy spectrum has a box
shape. For the upper and lower end-points $E_{+/-}$, the slepton
and LSP masses can be determined as
\begin{eqnarray*}
        m_{\tilde{\ell}} = 
        \frac{\sqrt{s}}{E_{-}+E_{+}}\,\sqrt{E_{-}\, E_{+}} & \quad &
        m_{\tchi^0_1}  =  m_{\tilde{\ell}} \,
          \sqrt{1 - \frac{E_{-}+E_{+}}{\sqrt{s}/2}}.
\end{eqnarray*}

The situation is more complicated in the case of $\tilde\tau$ sleptons
due to the escaping neutrinos from $\tau$ decay leading to a 
depletion of the of the upper end-point and to elimination of the
lower endpoint. However, if the mass of $\tchi^0_1$ is known e.g. from
smuon production, the shape and the upper end-point of the energy of the $\tau$
decay products can still be used to extract a precise mass value for
$\tilde\tau_1$ as shown for the $\tau\to 3\pi \nu_{\tau}$ spectrum in
Fig.~\ref{fig:boxes}(right).

\begin{figure*}[htbp]
\centering
\includegraphics[width=5cm]{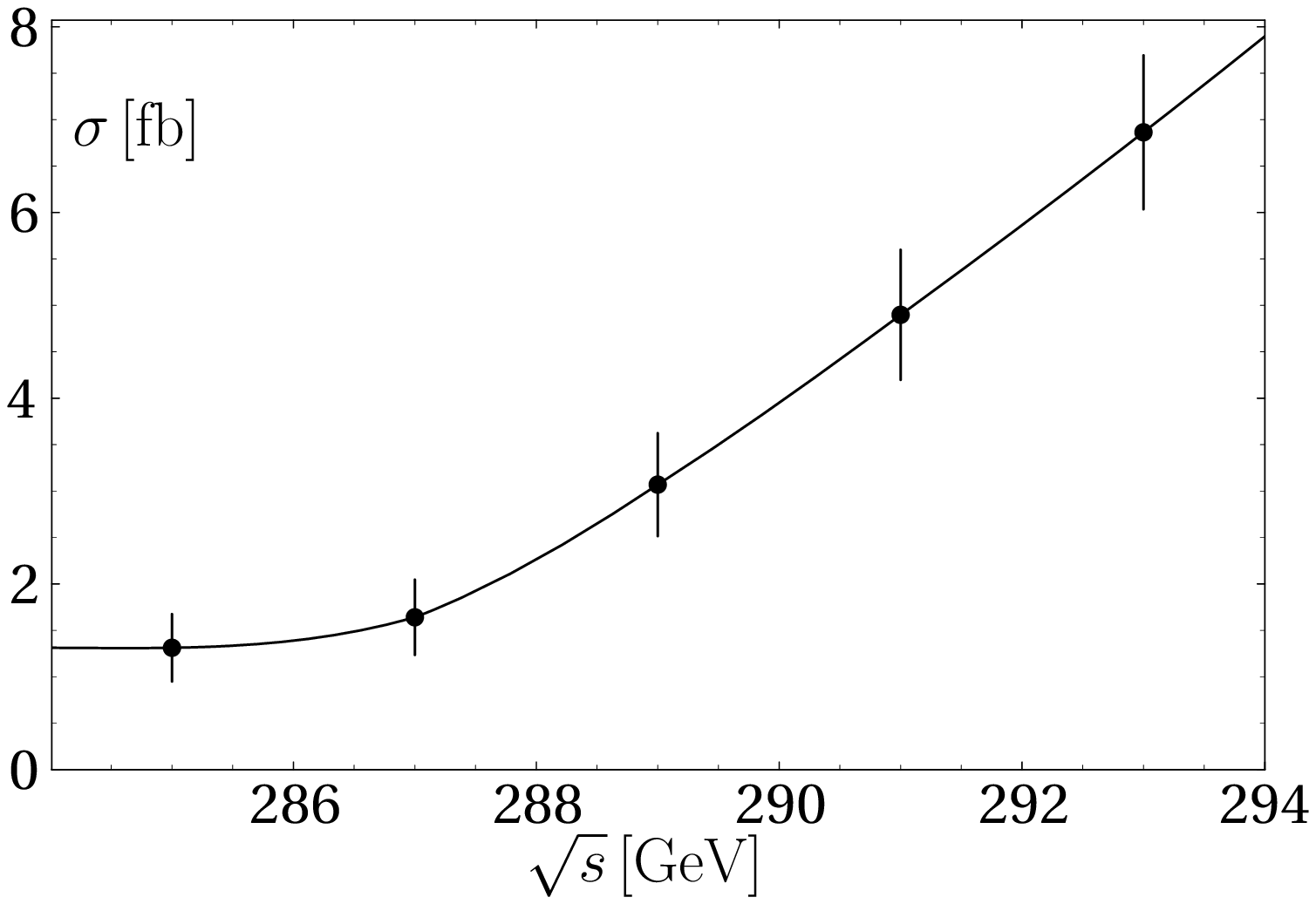}
\includegraphics[width=5cm]{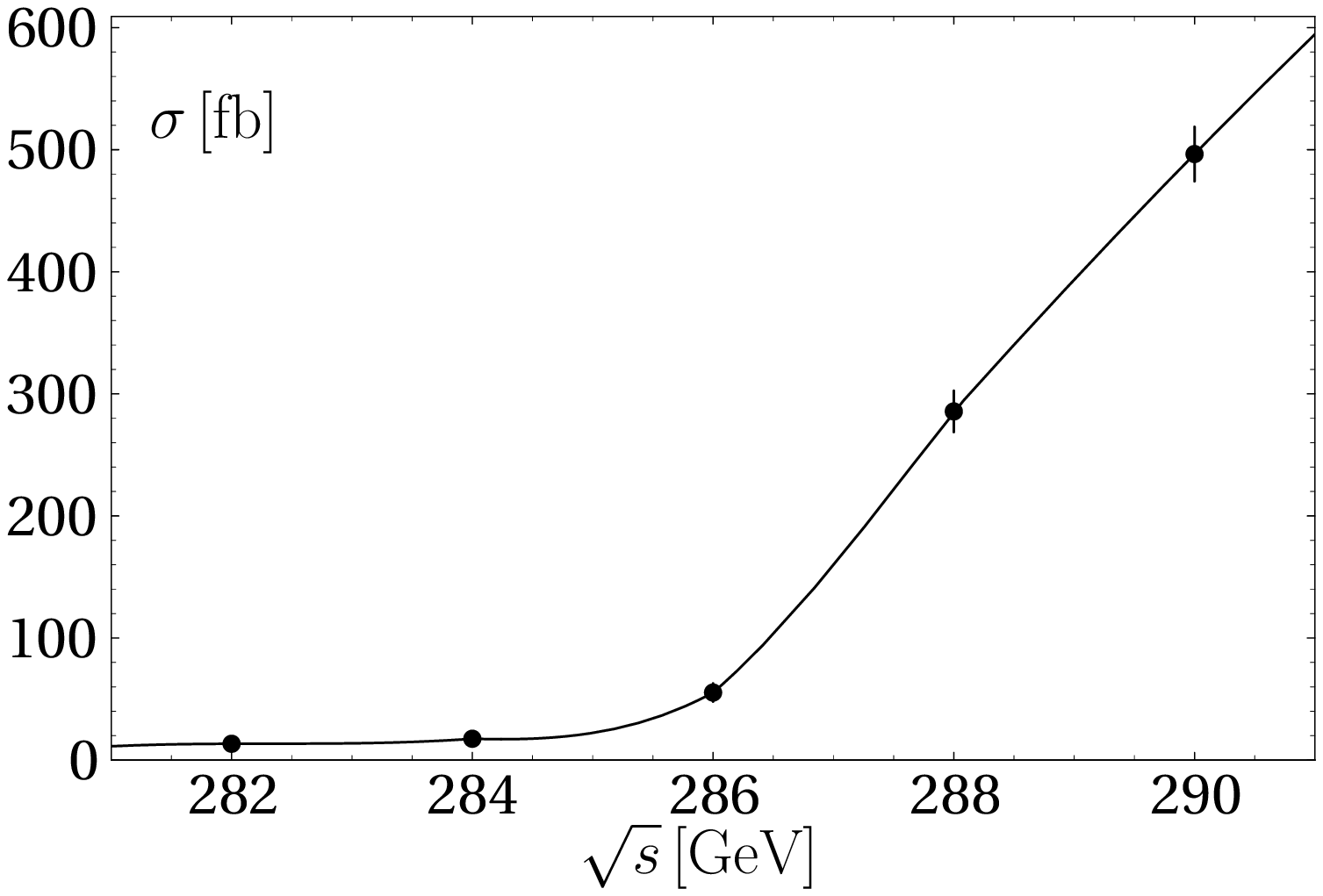}
\includegraphics[width=5cm]{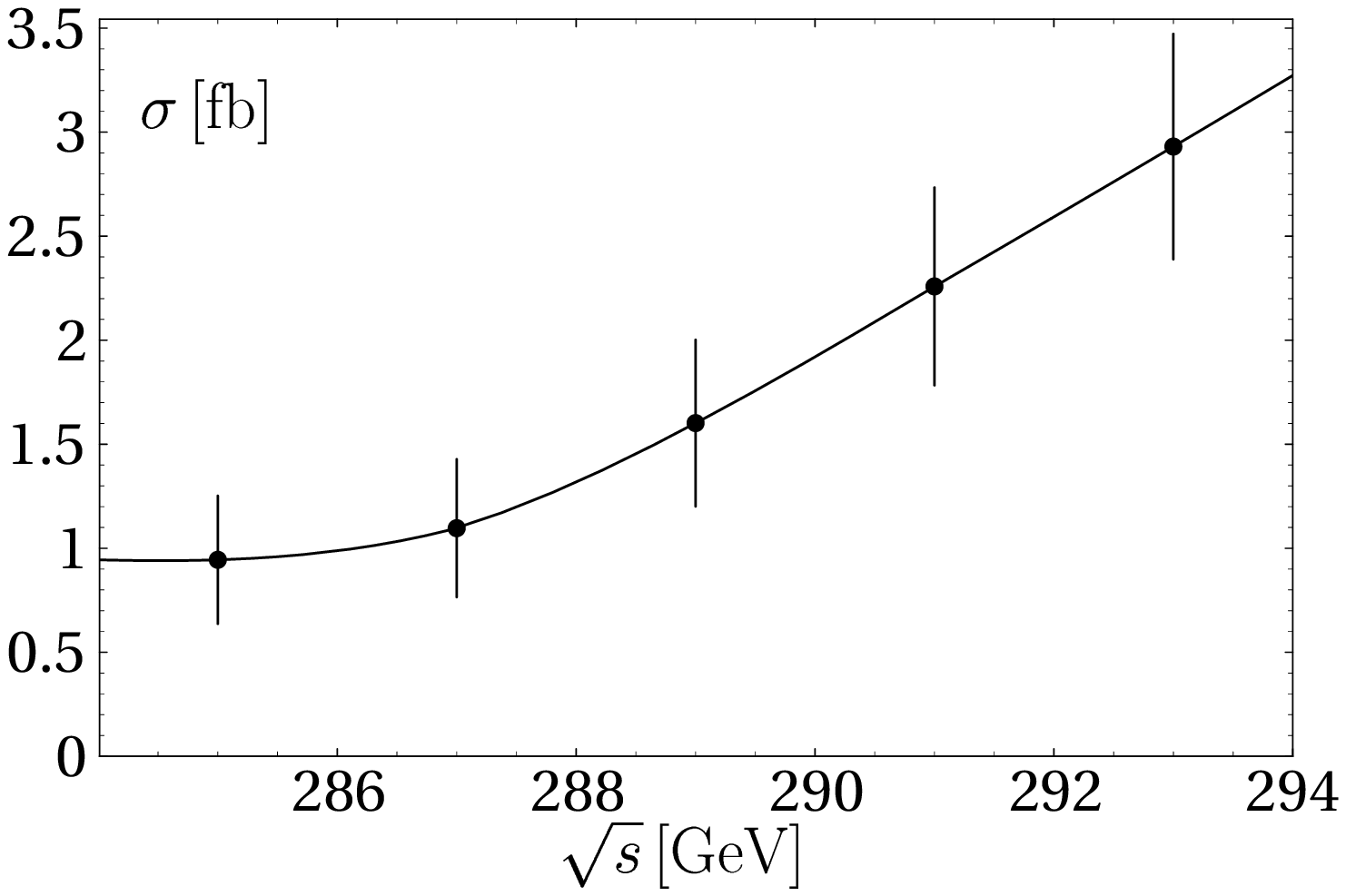}
\caption{cross-section for selectron and smuon pair production at threshold:
$e^+_L e^-_R\to\tilde{e}_R\tilde{e}_R$ (left),
$e^-_L e^-_R\to\tilde{e}_R\tilde{e}_R$ (middle),
$e^+_L e^-_R\to\tilde{\mu}_R\tilde{\mu}_R$ (right). The error bars
correspond to 10 fb$^{-1}$ per point.}
\label{fig:thresh}
\end{figure*}

Alternatively, the slepton masses can be extracted from a threshold
scan as shown in Fig.~\ref{fig:thresh} for right selectron production
both in $e^+e^-$ and $e^-e^-$ collisions and for right smuon production.
With measurements at five center-of-mass energies with only 10
fb$^{-1}$ per point a precision of $\cal O$(100 MeV) can be achieved.
With this precision higher-order corrections and final width
corrections have to be taken into account~\cite{freitas}.

\subsubsection{Charginos and Neutralinos}

Charginos and neutralinos are pair-produced
\begin{eqnarray}
    e^+e^- & \to & \tchi^\pm_i \tchi^\mp_j  \qquad \qquad \qquad [i,j = 1,2] \\
    e^+e^- & \to & \tchi^0_{i} \tchi^0_j  \qquad \qquad \qquad
                           [i,j = 1, \ldots ,4] 
\end{eqnarray}
via $s$-channel $\gamma/Z$ exchange and $t$-channel selectron or 
sneutrino exchange. 
The lightest chargino decays according to 
$\tchi^\pm_1 \to \ell^\pm \nu_\ell \tchi^0_1$ either
via an intermediate virtual or real $W^\pm$ boson or if kinematically
possible via a real slepton. 
The second lightest neutralino decays according to
$\tchi^0_2 \to \ell^+\ell^-\tchi^0_1$ either via a virtual or real
$Z$ boson or via a real slepton. In particular, if $m_{\tilde\nu} < 
m_{\tchi^0_2}$, invisible $\tilde{\nu}$ decays may occur.
For large mixing in the stau sector and for large values of
$\tan\beta$ the $\tilde\tau_1$ slepton is often much
lighter than the other sleptons which can lead to a significant
enhancement of $\tau$ leptons in the chargino and neutralino final
states. The production processes for $\tilde\tau$,
$\chi^0_2$ and $\chi^\pm_1$ may therefore all lead to the same
$\tau^+\tau^- + $ missing energy signature. Topological cuts and 
the use of polarized beams can help
to disentangle the contributing SUSY processes.
As in the case of sleptons, the chargino and neutralino masses can be
measured from the lepton energy and mass
spectra as well as from threshold scans. In the more difficult case of
exclusive decays into $\tau$ final states, a mass precision of a few
GeV can be achieved in the continuum and 0.5 GeV from a threshold scan.
Significantly better precision can be achieved if electron and muon
final states are produced with sufficient rate~\cite{charneu}.

\subsubsection{Light Stop}

Although squarks are often too heavy to be produced at a 1 TeV LC,
the light scalar top quark may be lighter than the other squarks and
therefore accessible in the reaction
$e^+e^- \to \tilde{t}_1\tilde{t}_1 \to b \tchi^+_1\,\bar{b} \tchi^-_1
        \to b \tau^+\nu\tchi^0_1 \, \bar{b} \tau^-\nu\tchi^0_1$.
The final state consists of two  $b$-jets, two $\tau$'s and missing
energy. The energy spectrum of the $b$-jets can be used to reconstruct
the stop mass provided the neutralino and chargino masses are known~\cite{feng}.
With a luminosity of 1000~fb$^{-1}$
the rate will be sufficient to achieve a mass resolution of 2 GeV.
For a light scalar top quark, the decay chain
$e^+e^- \to \tilde{t}_1\tilde{t}_1 \to 
        \to c \tchi^0_1 \, \bar{c} \tchi^0_1$ has also been studied.
From a measurement of the production cross-section with opposite
beam polarizations, a measurement of both mass and mixing angle can be
inferred~\cite{finch}.

\subsubsection{Masses: Summary}

The achievable superpartner mass precision of the ILC for the SPS1a
scenario is summarized in Table~\ref{tab:ilcmass} taken from~\cite{lhclc}.

\begin{table}[htbp]
\caption{Sparticle masses and their expected precisions in Linear
  Collider experiments, 
  SPS~1a mSUGRA scenario (from~\cite{lhclc}).}
\label{tab:ilcmass}

\centering
\begin{tabular}{|c|c|c|l|}
\hline
               & $m~[\GeV]$ & $\Delta m~[\GeV]$ & Comments\\ \hline
$\tilde{\chi}^\pm_1$ & 176.4           & 0.55      & simulation threshold scan ,
                                                     100 fb$^{-1}$ \\
$\tilde{\chi}^\pm_2$ & 378.2           & 3         & estimate
    $\tchi^\pm_1\tchi^\mp_2$, spectra $\tchi^\pm_2 \to Z \tchi^\pm_1,\, W \tchi^0_1$
               \\ 
\hline
$\tilde{\chi}^0_1$   &  96.1           & 0.05      & combination of all methods \\
$\tilde{\chi}^0_2$   & 176.8           & 1.2       & simulation threshold scan 
         $\tilde{\chi}^0_2\tilde{\chi}^0_2$,             100 fb$^{-1}$ \\
$\tilde{\chi}^0_3$   & 358.8           & 3 -- 5    & spectra
       $\tchi^0_3\to Z \tchi^0_{1,2}$, \ $\tchi^0_2\tchi^0_3, \tchi^0_3\tchi^0_4$, 750 GeV, $>1000~\fbi$ \\
$\tilde{\chi}^0_4$   & 377.8           & 3 -- 5    & spectra
      $\tchi^0_4\to W \tchi^\pm_1$, \  $\tchi^0_2\tchi^0_4, \tchi^0_3\tchi^0_4$,  750 GeV, $>1000~\fbi$ \\
\hline
$\tilde{e}_R$        & 143.0           & 0.05      & $e^-e^-$ threshold scan,
                                                     10 fb$^{-1}$ \\
$\tilde{e}_L$        & 202.1           & 0.2       & $e^-e^-$ threshold scan 
                                                     20 fb$^{-1}$ \\
$\tilde{\nu}_e$      & 186.0           & 1.2       & simulation 
                                                     energy spectrum, 500 GeV,
                                                     500 fb$^{-1}$ \\
$\tilde{\mu}_R$      & 143.0           & 0.2       & simulation
                                                     energy spectrum, 400 GeV,
                                                     200 fb$^{-1}$ \\
$\tilde{\mu}_L$      & 202.1           & 0.5       & estimate threshold scan,
                                                  100 fb$^{-1}$ \cite{grannis} \\
$\tilde{\tau}_1$     & 133.2           & 0.3       & simulation 
                                        energy spectra, 400 GeV, 200 fb$^{-1}$ \\
$\tilde{\tau}_2$     & 206.1           & 1.1       & estimate threshold scan,
                                            60 fb$^{-1}$  \cite{grannis} \\ 
\hline
$\tilde{t}_1$        & 379.1           & 2         & estimate 
                    $b$-jet spectrum, $m_{\rm min}(\tilde t)$, 1TeV, 1000 fb$^{-1}$ \\
\hline
\end{tabular} 

\end{table}

\subsection{Quantum numbers, couplings and mixings}

Besides the precise measurement of the largest possible set of
superpartner masses the measurement of quantum numbers, couplings,
and mixings plays an important role in deciphering the supersymmetric model.
In $e^+e^-$ collisions, due to the low background and the known initial
state, various possibilities to extract quantum numbers and couplings
exist. These range from the measurement of inclusive rates to the
measurement of angular distributions in production and decay.

\subsubsection{Spin determination}

\begin{figure*}[htbp]
\centering
\includegraphics[width=7cm]{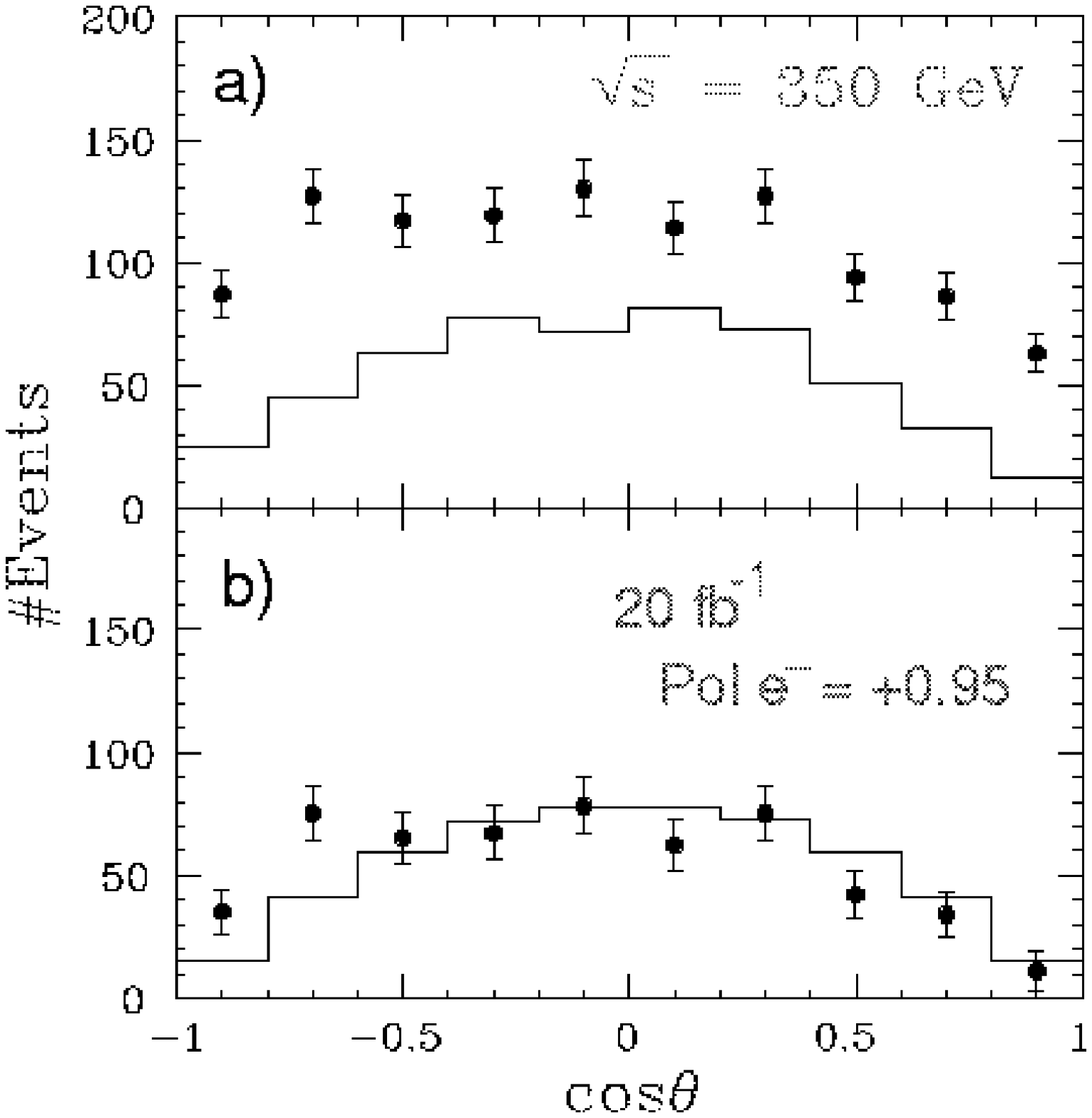}\quad\quad
\includegraphics[width=7cm]{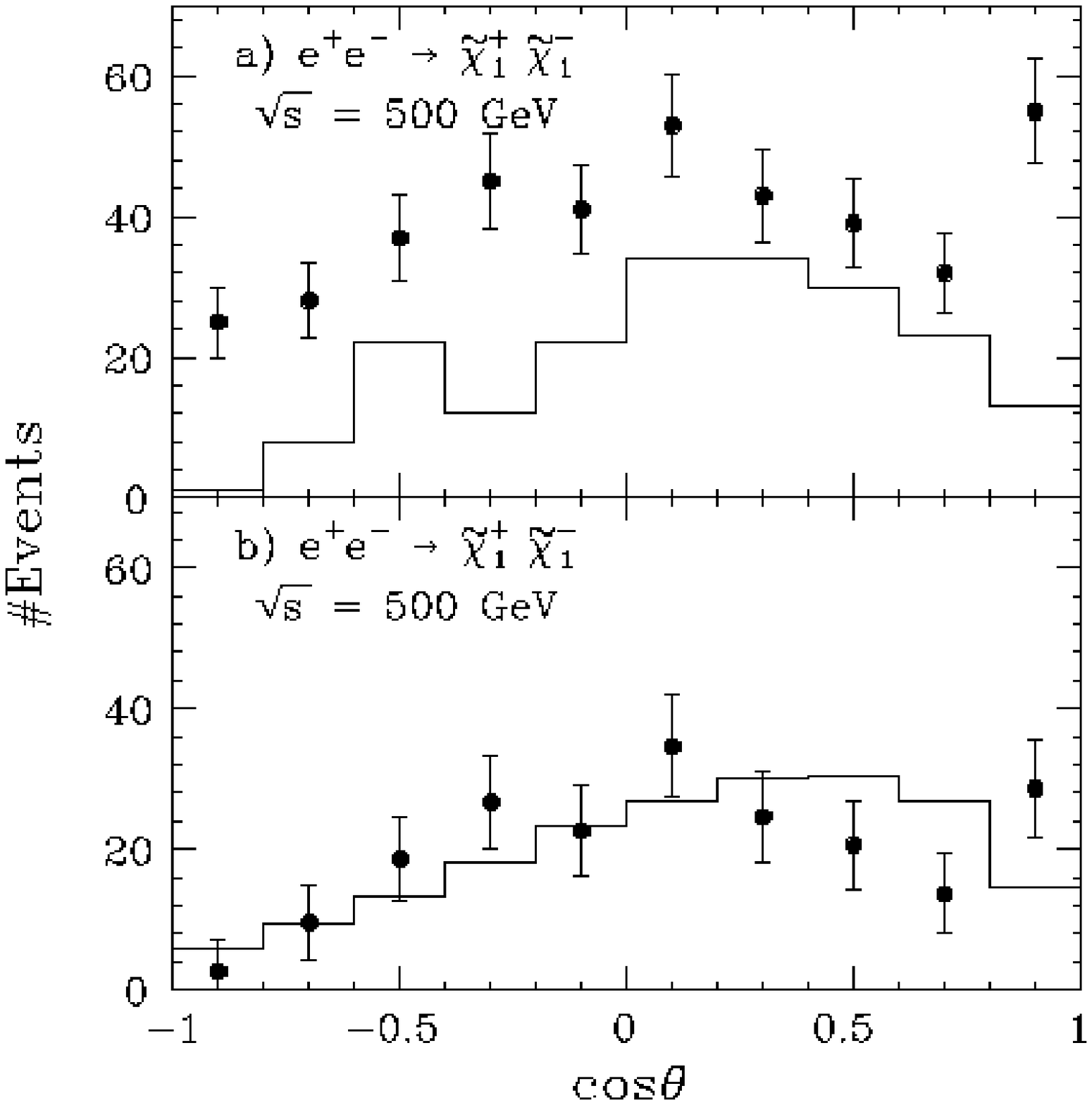}
\caption{Distribution of the superpartner production angle for
smuon (left) and charginos (right). The upper plots show both
reconstructed solutions per event as points and the correct
solution as full line; in the lower plots the combinatoric background
is subtracted (from~\cite{acfareport}).}
\label{fig:costh}
\end{figure*}

The fundaments of SUSY rely on the superpartners' spin differing by
$\frac{1}{2}$ from their SM partners. It was shown in Sec.~\ref{sec:lhcspin}
that at the LHC a unique spin determination is quite involved. At the LC,
the spins of the superpartners can be determined directly from the 
production angle distributions. The scalar leptons exhibit a $\sin^2\theta$
distribution which can be reconstructed up to a twofold ambiguity 
in smuon pair-production. After subtraction of the combinatorial background
the angular distribution is very 
clean~(Fig.~\ref{fig:costh}, left)~\cite{acfareport}.
The situation is more complicated for charginos and neutralinos which
exhibit a forward-backward asymmetry in the production angle due to their
mixed U(1) and SU(2) couplings and the additional t-channel contribution.
An example for chargino pair production is shown in 
Fig.~\ref{fig:costh}, right). The forward-backward asymmetry and in particular
the left-right polarization asymmetry provide sensitive observables in order
to disentangle the chargino and neutralino mixing matrices~\cite{gudiafb}.

\subsubsection{Chiral quantum numbers}

\begin{figure*}[htbp]
\begin{center}
     \vspace*{5.8cm}
\begin{picture}(7,7)
\setlength{\unitlength}{1cm}
\put(-8,0){\mbox{\includegraphics[height=.25\textheight]{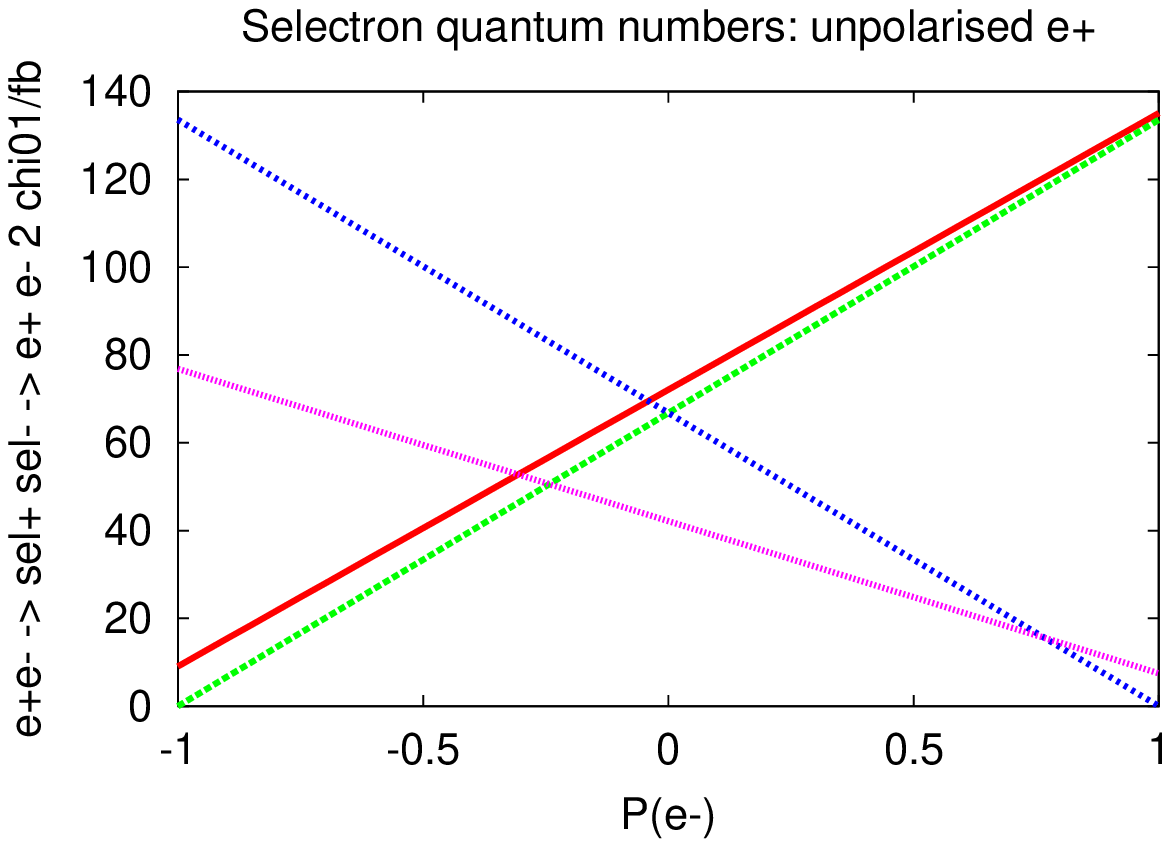}}}
\put(-1.5,3.1){\small \makebox(0,0)[br]{{\color{Green}
$\tilde{e}^+_{L} \tilde{e}^-_{R}$}}}
\put(-3.1,1.6){\small \makebox(0,0)[br]{{\color{Lila}
$\tilde{e}^+_{L} \tilde{e}^-_{L}$}}}
\put(-4.1,3.7){\small \makebox(0,0)[br]{{\color{Blue}
$\tilde{e}^+_{R} \tilde{e}^-_{L}$}}}
\put(-1.9,4.0){\small \makebox(0,0)[br]{{\color{Red}$
\tilde{e}^+_{R} \tilde{e}^-_{R}$}}}
\put(-6,4.7){\tiny $\sqrt{s}=500$~GeV}

\put(.5,0){\mbox{\includegraphics[height=.25\textheight]{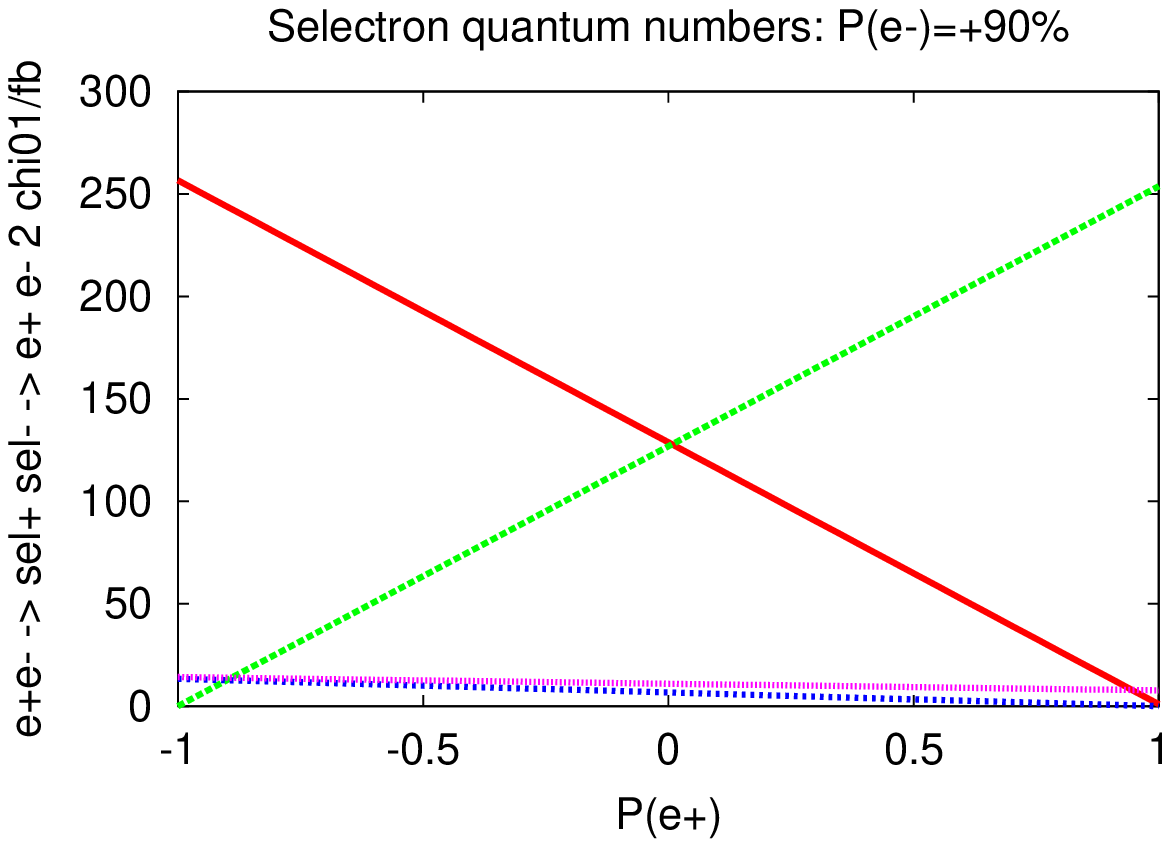}}}
\put(6.8,2.7){\small \makebox(0,0)[br]{{\color{Green}
$\tilde{e}^+_{L} \tilde{e}^-_{R}$}}}
\put(3.5,3.1){\small \makebox(0,0)[br]{{\color{Red}$
\tilde{e}^+_{R} \tilde{e}^-_{R}$}}}
\put(2.4,4.7){\tiny $\sqrt{s}=500$~GeV}
\end{picture}\vspace{1cm}
\end{center}\vspace{-1.5cm}
\caption{\label{fig:selpol} Separation of the selectron pair
$\tilde{e}_L^-\tilde{e}^+_R$ in $e^+ e^-\to \tilde{e}^+_{L,R}
\tilde{e}^{-}_{L,R}$ may not be possible with 
electron polarization only (left); if, however, both beams are polarized, the
$RR$ configuration separates the pairs and the association of the selectrons to the chiral 
quantum numbers can be experimentally tested (from \cite{gudipol}).}
\end{figure*}

In SUSY, the chiral (anti-)fermions are associated in an unambiguous way
to scalars, i.e.~$e^-_{L,R} \leftrightarrow \tilde{e}^-_{L,R}$ and
$e^+_{L,R} \leftrightarrow \tilde{e}^+_{R,L}$. The four pair-production 
processes for left and right selectrons, 
$e^+e^-\to \tilde{e}^+_R\tilde{e}^-_R$,
$e^+e^-\to \tilde{e}^+_L\tilde{e}^-_L$,
$e^+e^-\to \tilde{e}^+_R\tilde{e}^-_L$,
$e^+e^-\to \tilde{e}^+_L\tilde{e}^-_R$ can be disentangled from their
different dependence of the cross-section to polarized electron and
positron beams. From Fig.~\ref{fig:selpol}, it can be seen that 
e.g.~$e^+e^-\to \tilde{e}^+_R\tilde{e}^-_R$ and 
$e^+e^-\to \tilde{e}^+_L\tilde{e}^-_R$ have practically identical behavior
of the cross-section as a function of the electron polarization but differ
completely as a function of the positron polarization~\cite{gudipol}.
The t-channel contribution to the production cross-sections is sensitive
to the SUSY Yukawa coupling $\hat{g}(e\tilde{e}\tchi^0)$ which is fundamentally
related to the SM gauge couplings. The SU(2) and U(1) SUSY Yukawa
couplings can be determined to a precision of 0.7\% and 0.2\%, 
respectively with 500 fb$^{-1}$ at 500~GeV in a SPS1a scenario~\cite{sleptons}.

\subsubsection{Stau mixing angle}

In the third generation large mixing effects are expected. Again, beam 
polarization is vital to measure the stau mixing angle from measuring
the cross-section for $e^+e^-\to\tilde{\tau}_1\tilde{\tau}_1$ with
polarized beams. This allows for a measurement of 
$\cos{2 \theta_{\tilde{\tau}}} = -0.84 \pm 0.04$ in an SPS1a 
scenario for 500~fb$^{-1}$~\cite{staumix}. 
Further, independent information can be obtained
from the measurement of the $\tau$ lepton polarization in 
$\tilde{\tau}_1\to\tau\tchi^0_1\to\pi^\pm\nu_\tau\tchi^0_1$.
For known $\tilde\tau$ mixing angle, this information can be used to
gain sensitivity to $\tan\beta$. 

\subsubsection{CP-violation in SUSY decays}

Many studies of SUSY at LHC and LC have focused on a (often constrained) 
MSSM with real parameters. The general MSSM Lagrangian however allows for
CP-violating complex parameters. As an example, the U(1) gaugino mass parameter
$M_1$ and the Higgsino mixing parameter $\mu$ may have complex phases
$\phi_{M_1}, \phi_{\mu}$. These phases may be accessed in angular correlations
of the decay products in neutralino decay~\cite{bartl}. An experimental 
study of how well these correlations can be measured is still lacking.

Similarly, CP-violating phases of the tri-linear couplings $A_\tau, A_b, A_t$
may be accessed in studying the branching fractions of the third generation
sfermions an analyze them together with their masses and production cross
sections in a global fit~\cite{hesselbach}.

\subsection{Constraining Dark Matter}

The SUSY LSP provides an excellent candidate for dark matter. Recent 
measurements of temperature fluctuations of the cosmic microwave background
by the WMAP satellite~\cite{wmap} 
strongly constrain the SUSY LSP properties and therefore point to certain
regions in the MSSM parameter space. Of particular interest for experimental
studies at colliders is the co-annihilation region in which the
neutralino annihilation is enhanced by the t-channel process 
$\tchi\tilde{\tau}\to\tau\gamma$ which contributes significantly only if
the mass difference 
$\Delta m = m(\tilde\tau)-m(\tchi^0_1)$ is small. The relic 
dark matter density depends critically on this mass difference. With the
next generation of CMB experiments, in particular Planck, the DM density
can be measured at the 2-3\% level. It is therefore imperative to match this
precision at colliders. 

\begin{figure*}[htbp]
\centering
\includegraphics[width=15cm]{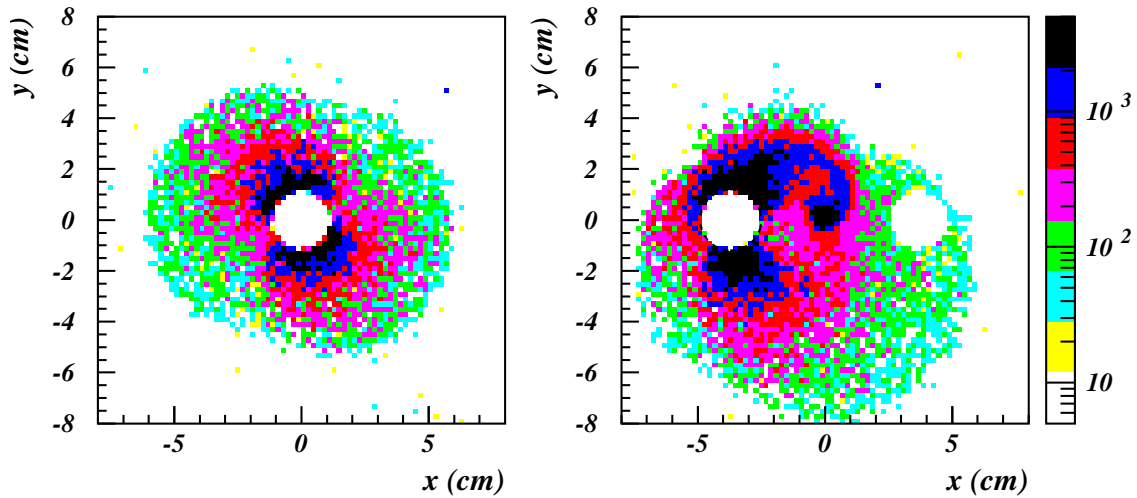}
\caption{Energy deposition from beam-strahlung (in GeV) in the forward
detector region (3.7m from the interaction point) 
without (left) and with (right) a 10 mrad half-angle 
crossing angle. }
\label{fig:maskbkg}
\end{figure*}

If $\Delta m$ is small (typically below 10~GeV), 
the staus decay with small visible energy
and the signature is only a few soft charged tracks accompanied by large 
missing energy. Two-photon background is becoming severe unless it can
be efficiently vetoed by the detection of very forward scattered electrons.
In the very forward region significant energy induced by beam-beam-interactions
is deposited. This energy deposition in the most forward calorimeter, 
is shown in Fig.~\ref{fig:maskbkg} for two interaction region
designs without (left) and with (right) a crossing-angle of the two incoming
beams. The detection of energetic electrons is possible down to angles
of 3.5 (5.7) mrad without (with) crossing angle if the calorimeter is finely
segmented. 

The pair production of staus in the small-$\Delta m$ region
has been studied in~\cite{martyn-smalldm} and~\cite{bambade} for various
MSSM parameter sets. With appropriate cuts, detection and a precise measurement
of the $\tilde\tau$ mass is possible down to $\Delta m$
$\sim$3~GeV. The resulting
precision on the prediction for the dark matter density ranges from 2~to 6\%,
depending on the $\tilde\tau$ mass and on $\Delta m$. This precision matches
the anticipated precision of the Planck satellite of 2\%. As an example
the hadronic energy spectra for $\tau$ decays from the process
$e^+_L e^-_R\to \stau_1  \, \stau_1 \to \tau^+ \tchi^0_1 \,\, 
\tau^-\tchi^0_1 $ as shown after detector simulation and cuts together
with the two-photon background for $\Delta m = 5 $ GeV (Model Point D$^\prime$
from~\cite{battagliabench}).


\begin{figure}[htbp]
\includegraphics[height=15cm,angle=90]{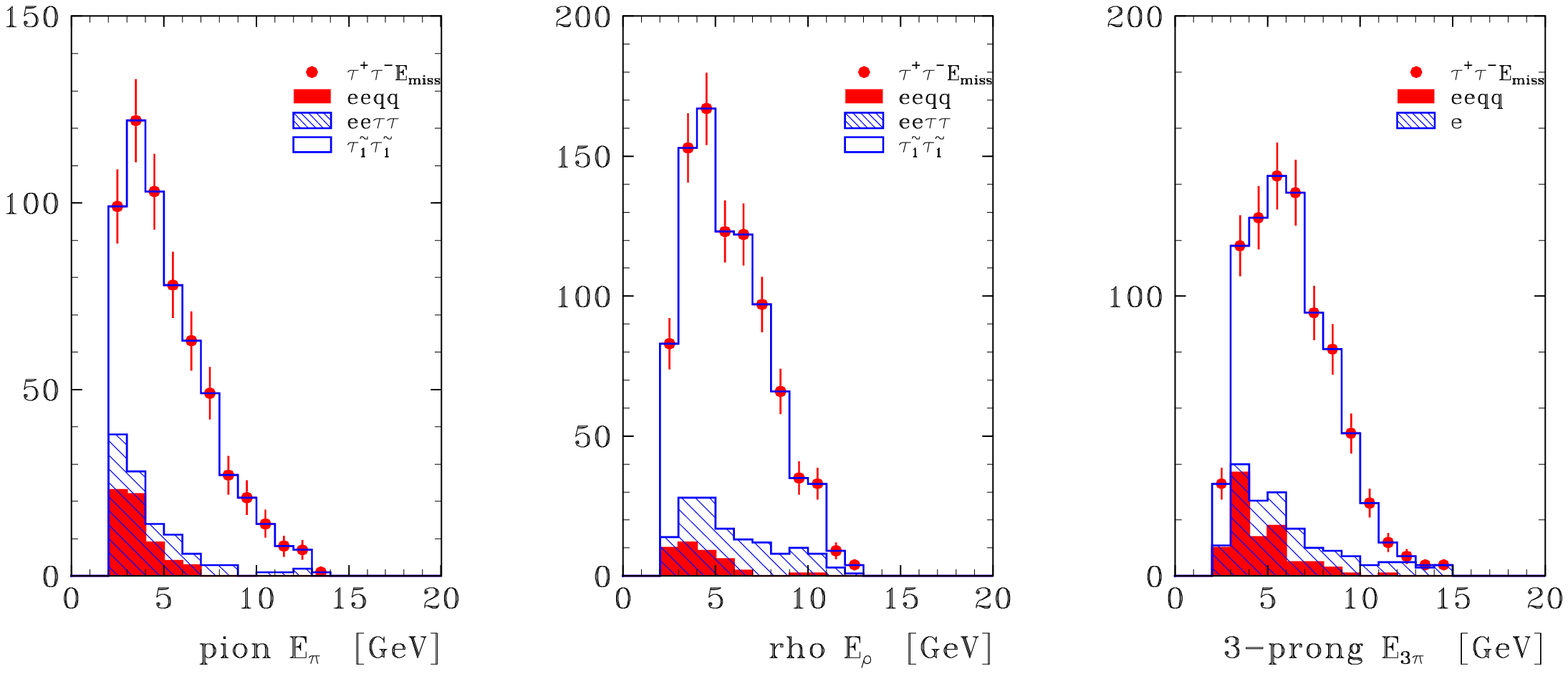}

  \caption{Hadron energy spectra $E_\pi$ of $\tau\to\pi\nu_\tau$,
    $E_\rho$ of $\tau\to\rho\nu_\tau$ and
    $E_{3\pi}$ of $\tau\to 3\pi\nu_\tau$ decays
    from the reaction
    $e^+_L e^-_R\to \stau_1  \, \stau_1   
    \to \tau^+ \tchi^0_1 \,\, \tau^-\tchi^0_1 $ and 
    two-photon production
    assuming head-on collision.
    Model D', $m_{\stau_1} = 217.5\;\GeV, \ \Delta m = 5.1\;\GeV$,
    $\sqrt{s}=600\;\GeV$ and $\cL=300\;\fbi$}
  \label{etaud_spectrum}
\end{figure}

\section{LHC/LC INTERPLAY}
\label{sec:lhclc}

The ultimate goal of measurements of the properties of superpartners at
LHC and LC will be the extraction of the complete set of parameters of the low
energy MSSM Lagrangian. The previous sections have already indicated the
complementarity of the possibilities at the LHC (large mass reach for
squarks and gluinos) and the LC (precision measurements of color-neutral
part of spectrum). In this section, we will discuss the additional benefit
of simultaneous interpretation and possibly simultaneous data analysis
if both machines run concurrently. This aspect has recently been studied
in the international LHC/LC working group~\cite{lhclc}. Without aiming for
completeness, a few examples of this interplay will be given in the 
following.

\subsection{Joint analysis of superpartner masses}

\begin{figure}[htbp]
\includegraphics[width=7cm]{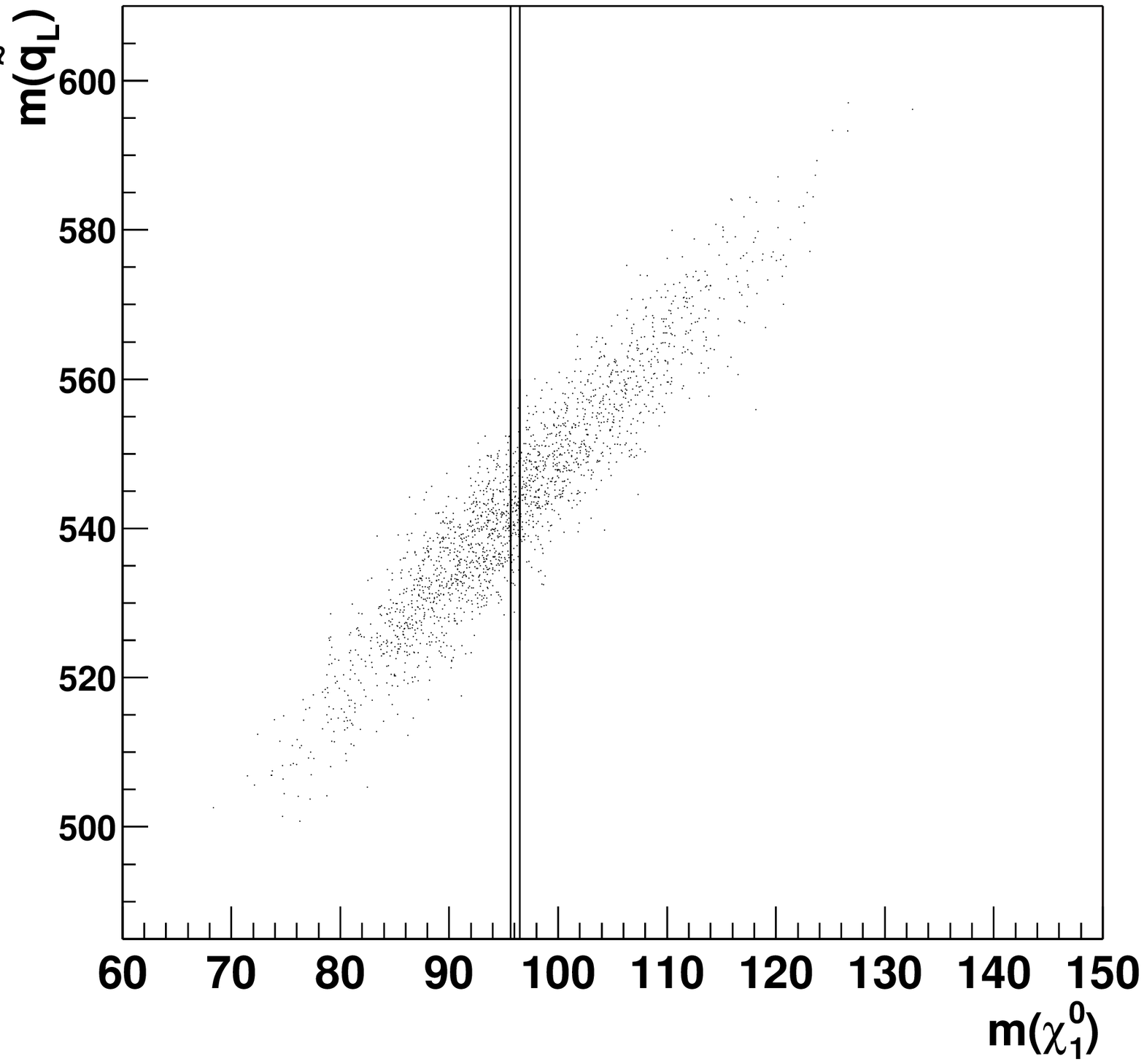}\quad\quad
\includegraphics[width=7cm]{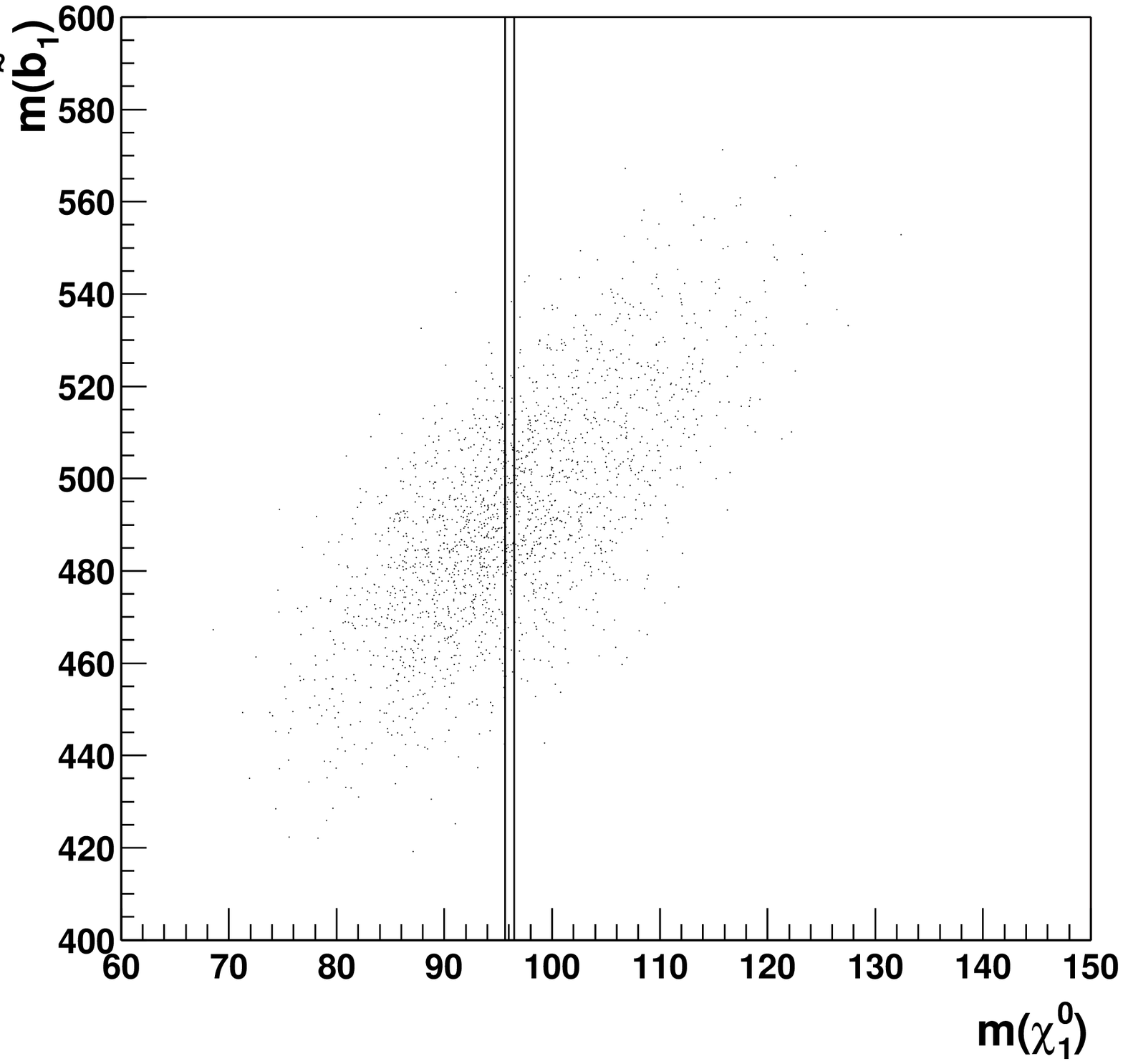}
\caption{Correlation between left squark (left) and lighter sbottom
(right) and the LSP mass at the LHC.}
\label{fig:lhccorr}
\end{figure}

It has been shown in Section~\ref{sec:lhc} that mass reconstruction
of squarks and gluinos at the LHC suffers from the unknown masses of the
lighter states, in particular the LSP and the sleptons. 
Under favourable circumstances, they can be reconstructed
from a joint fit of various kinematic endpoints to moderate accuracy. 
However a strong correlation of e.g.~the squark mass and the LSP neutralino
mass remains (see Fig.~\ref{fig:lhccorr}). A precise measurement of
slepton, chargino, and neutralino masses at the LC removes these correlations
and improves the precision of squark and gluino masses considerably even
in a situation where the latter are not directly accessible at the LC.
The expected precisions on some of the masses in a SPS1a scenario 
are shown Tab.~\ref{tab:susymasseslhclc}. It should be noted that the
precision on squark and gluino masses in the combined analysis is dominated
by uncertainties on the hadronic energy scale of the LHC experiments,
assumed to be 1\%.

\begin{table}[htp]
\begin{center}
\caption{The RMS values of the mass distribution in the case of 
the LHC alone, and together with the ILC
measurement of $m_{\tilde{\chi}_1^0}$.
Only the masses resulting from the edge analyses
are given. All numbers in GeV.} \vspace{4pt}
\label{tab:susymasseslhclc}
\begin{tabular}{ccc}
\hline\hline
 & LHC & LHC+ILC \\
\hline
$\Delta m_{\tilde{\chi}_1^0}$& 4.8 & 0.19\\
$\Delta m_{\tilde{l}_R}$ & 4.8 & 0.34\\
$\Delta m_{\tilde{\chi}_2^0}$& 4.7 & 0.24\\
$\Delta m_{\tilde{q}_L}$ & 8.7  & 4.9\\
$\Delta m_{\tilde{b}_1}$ & 13.2 & 10.5\\
\hline
\end{tabular}
\end{center}
\end{table}

\subsection{ILC mass predictions for LHC searches}

At the ILC, the SUSY parameters which govern the chargino-neutralino sector,
i.e. the U(1) and SU(2) gaugino mass parameters $M_1, M_2$, the Higgsino
mixing parameter $\mu$ and $\tan\beta$ can be uniquely and precisely
extracted from the measurements of masses and polarized cross-sections
of the lightest gauginos, $\tchi^0_1, \tchi^0_2, $ and $\tchi^\pm_1$.
These parameters can in turn be used to predict the masses of the heavier
members $\tchi^0_3, \tchi^0_4$ and $\tchi^\pm_2$ within the framework of
the general MSSM. With such a mass prediction, the search for the (often
small) signals of the heavy gauginos can be substantially facilitated.
The search for an edge in the di-lepton mass spectrum becomes transformed
into a single hypothesis test, thus inceasing the statistical power of
the LHC data for this specific hypothesis. 

Should in turn the predicted state be observed at the LHC, its measured
mass can be fed back into the SUSY parameter analysis and considerably improve
the achieavble precision.

Furthermore, the comparison of the predicted $\tchi^0_4$ mass from LC
measurements and the measured mass at the LHC allows to test for models
beyond the MSSM. This has recently been shown~\cite{gudinmssm} for a 
NMSSM, where in contrast to the MSSM five neutralinos are predicted with
a mass spectrum not satisfying the MSSM relations. 


\subsection{Global MSSM fits and reconstruction of the theory}

Ultimately, the goal is to measure the complete set of electro-weak
scale parameters of the SUSY Lagrangian. At tree level, the parameter 
determination can proceed sector by sector, e.g.~the chargino sector
is completely determined by the three parameters $M_2, \mu,
\tan\beta$. However, with the anticipated precision of ILC
measurements, higher order corrections to masses, cross-sections, and 
branching ratios are not negligible. At loop-level in principle every
observable depends on the full set of SUSY parameters. An analytic
procedure to extract the Lagrangian parameters from data is 
no longer possible. Instead, a global fit of the Lagrangian parameters
to the complete set of SUSY observabales at LHC and LC will be
necessary.

Two programs, SFITTER~\cite{sfitter} and Fittino~\cite{fittino} have
been developed to achieve this goal. As an example, a recent result
from Fittino is explained here. A MSSM with 19 free parameters has
been chosen as the theoretical basis. It is derived from the full
MSSM but assuming real couplings, flavour diagonal sfermion mass matrices
and universality of the soft SUSY
breaking parameters of the first two generations.
For the theoretical predictions for the observables as a function of the
parameters the SPHENO~\cite{spheno} program has been used. This code
includes higher-order corrections whereever they have been calculated.
The program is interfaced via the 'SUSY Les Houches Accord' 
(SLHA)~\cite{slha}, a convention to exchange SUSY parameter
information between various programs in a coherent way. Therefore it
is possible to replace the actual SUSY code behind Fittino in order to
perform comparisons.
As simulated measurements, the masses measureable at LHC and LC as
explained in the previous sections have been used. In addition
polarized cross-section measurements at 500 and 1000 GeV LC have been
input with accuracies based on estimates from the expected
statistical errors, but including systematic errors of 1\% in each
case. For the true values of the parameters, the SPS1a scenario has
been chosen but nowhere in the fit any assumptions on the particular
SUSY breaking mechanism have been made. Special care was given to a
fitting strategy which does not make use of any a-priori knowledge of
the parameters. In particular, suitable start values for the
parameters have been estimated from the measurements using tree-level
relations between various observable and parameters sector by sector.
In order to yield a correctly converging fit an iterative procedure
has to be applied which does not leave free all parameters from the
beginning. Only after this preparatory phase the full fit can be
performed with all parameters left free. The example result for SPS1a is
shown in Table~\ref{tab:fittino}. The importance of higher-order
corrections can be seen from comparing the tree-level parameter
estimates with their final values. It has been verified
that the obtained fit errors are consistent with the fluctation of the
fitted parameters in many repeated measurements. It should be noted that
neither LHC nor LC input alone can constrain the assumed model enough
to yield a converging fit. It was also shown, that due to the
mutual influence of many parameters in determining a single
observable, wrong assumptions of un-fitted parameters lead to wrong
central values for the fit parameters. The lesson from this
excercise are: first, it is possible, at least for a favourable
scenario like SPS1a, to extract the complete electro-weak scale
Lagrangian from the future measurements of LHC and LC without strong
assumptions on the particular SUSY breaking scenario. Second, the fact
that results from both LHC and LC are necessary to obtain this result
nicely shows the strong interplay between both machines. Third, 
the achievable experimental precision clearly requires the knowledge
of theoretical predictions beyond leading order. The definition of a
clear scheme to extract well-defined parameters at higher orders is
currently being worked out in the Supersymmetry Parameter Analysis
(SPA) project~\cite{spa}.

\begin{figure}[htbp]
\includegraphics[width=5cm]{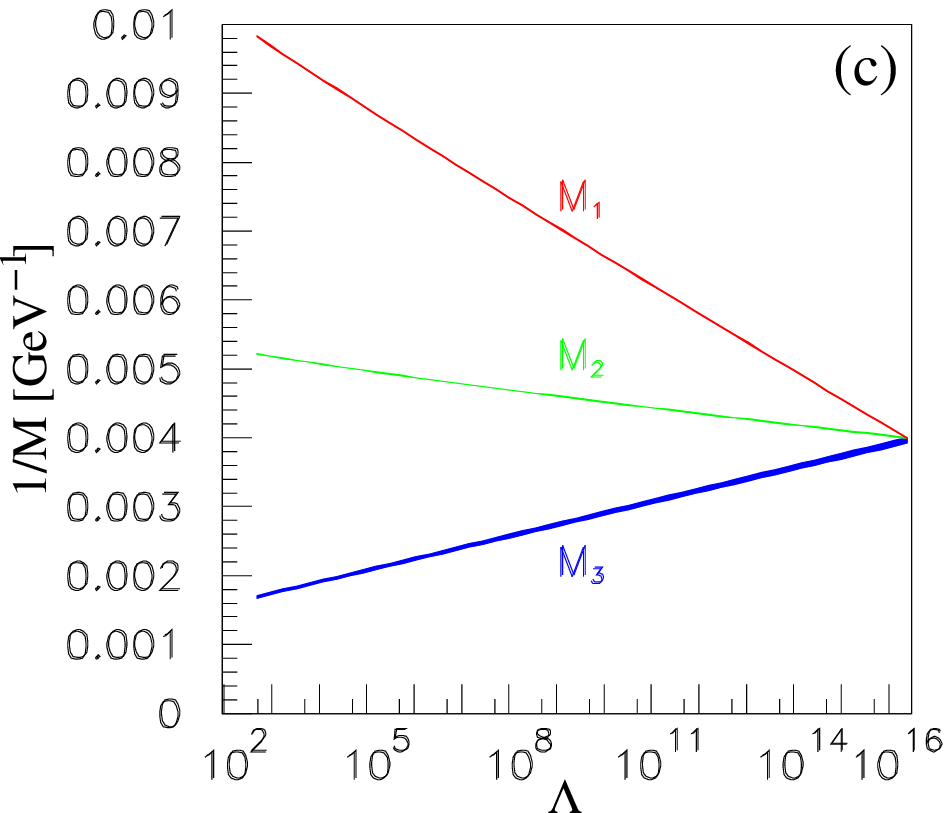}\quad
\includegraphics[width=5cm]{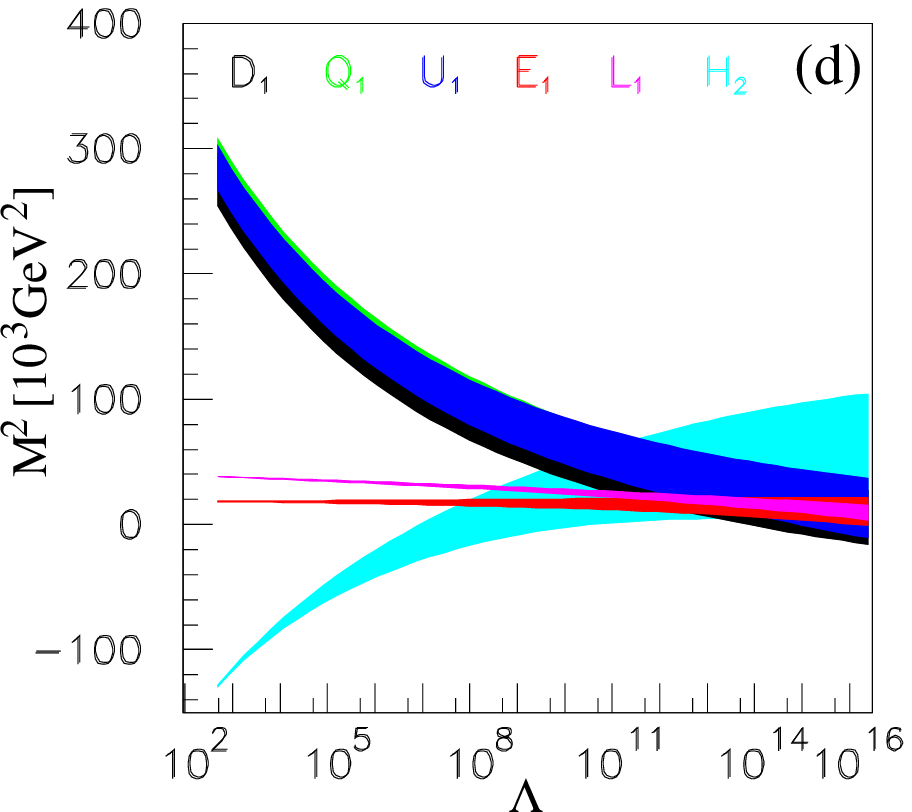}\quad
\includegraphics[width=5cm]{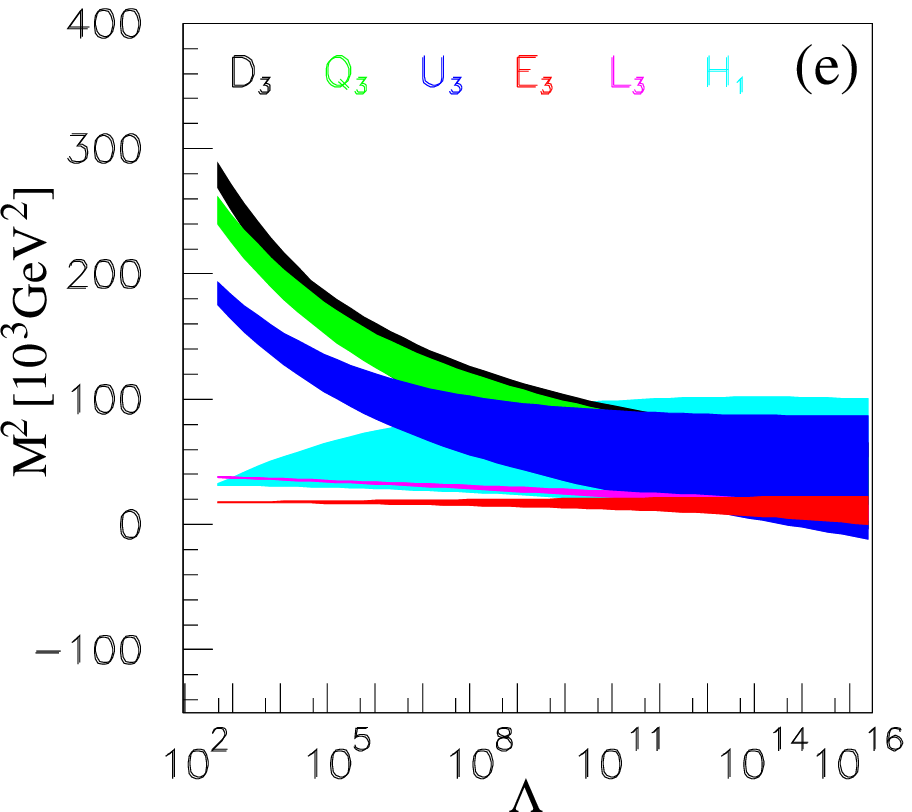}

\caption{Renormalization group evolution of the MSSM parameters 
determined from LHC and LC measurements to the GUT scale.}
\label{fig:gut}
\end{figure}

The extracted parameters of the electro-weak scale MSSM Lagrangian can
then be extrapolated to high (GUT, Planck) scales (Fig.~\ref{fig:gut}
in order to
determine distinct patterns of unification and reconstruct the
underlying fundamental theory of SUSY breaking~\cite{porod}.

\begin{table}[t]
\caption[The Fittino SPS1a fit result]{\sl The Fittino SPS1a fit result. 
  The left column shows the assumed SPS1a values, the middle column represents the 
  result of the intermediate fit without $m_{\mathrm{top}}$, and the right column 
  shows the result of the final fit. All SPS1a input values of the parameters are 
  reconstructed.}\label{tab:fittino}
\begin{center}
\begin{tabular}{|l|c|c|c|}
\hline
Parameter & SPS1a value & Tree-level &  Final fit result  \\
          &                              & estimate   &                  \\
 \hline\hline
 $\tan\beta$               &  10.0         &  9.97 &       $    10.0 \pm       0.3 $  \\
   $\mu$               &  358.64 GeV   &  354.4 GeV &      $    358.6 \pm      1.1 $ GeV  \\
$X_{\tau}$             &  -3837.23 GeV &  -3533.0 GeV &    $   -3837.2 \pm    131.0$ GeV  \\
$M_{\tilde{e}_R}$      &  135.76 GeV   &  150.2 GeV &      $    135.76 \pm    0.39 $ GeV  \\
$M_{\tilde{\tau}_R}$   &  133.33 GeV   &  141.0 GeV &      $    133.33 \pm    0.75 $ GeV  \\
$M_{\tilde{e}_L}$      &  195.21 GeV   &  202.7 GeV &     $     195.21 \pm   0.18 $ GeV  \\
$M_{\tilde{\tau}_L}$   &  194.39 GeV   &  206.6 GeV &     $     194.4 \pm     1.18 $ GeV    \\
$X_{\mathrm{top}}$     &  -506.388 GeV &  -43.5 GeV &     $   -506.4 \pm      29.5 $ GeV      \\
$X_{\mathrm{bottom}}$  & -4441.0 GeV    & -3533.0 GeV &   $   -4441.1 \pm   1765 $ GeV  \\
$M_{\tilde{d}_R}$      &  528.14 GeV   &  567.3 GeV &      $    528.2 \pm      17.6 $ GeV  \\
$M_{\tilde{b}_R}$      &  524.718 GeV  &  566.0 GeV &     $    524.7 \pm      7.7 $ GeV  \\
$M_{\tilde{u}_R}$      &  530.253 GeV  &  566.9 GeV &     $    530.2 \pm      19.1 $ GeV  \\
$M_{\tilde{t}_R}$      &  424.382 GeV  &  373.7 GeV &     $    424.4 \pm      8.54 $ GeV  \\
$M_{\tilde{u}_L}$      &  548.705 GeV  &  581.3 GeV &       $    548.7 \pm      5.2 $ GeV  \\
$M_{\tilde{t}_L}$      &  499.972 GeV  &  575.4 GeV &       $    500.0 \pm      8.1 $ GeV  \\
      $M_1$            &  101.809 GeV  &  99.07 GeV &     $    101.81 \pm    0.06 $ GeV  \\
      $M_2$            &  191.7556 GeV &  195.08 GeV&      $    191.76 \pm    0.10 $ GeV  \\
      $M_3$            &  588.797 GeV  &  630.5 GeV &     $    588.8 \pm      7.9 $ GeV  \\
  $m_A$                &  399.767 GeV  &  399.8 GeV &      $    399.8 \pm      0.71 $ GeV  \\
 $m_{\mathrm{top}}$    &  174.3 GeV    &  174.3 GeV  &      $      174.3 \pm  0.34 $ GeV  \\
\hline
\end{tabular}
\end{center}
\end{table}

\begin{acknowledgments}

The author would like to thank SLAC for the invitation to speak
the SLAC Summer Institute 2004 and the organizers for their
great hospitality. He also likes to thank H.-U.~Martyn and G.~Polesello 
for their careful reading of the manuscript and for many useful comments.
\end{acknowledgments}



\end{document}